\documentclass[10pt,aps,prd,twocolumn,preprintnumbers,superscriptaddress,nofootinbib,floatfix,notitlepage]{revtex4-2}

\usepackage{axodraw2,pix}
\usepackage{epsfig}
\usepackage{amsfonts}
\usepackage{amsmath}
\usepackage{bbm,bm}
\usepackage[dvips]{xcolor}
\usepackage{graphicx}
\usepackage[linkbordercolor={red!80!black},citebordercolor={1 1 1},urlbordercolor={1 1 1}]{hyperref}
\usepackage{orcidlink}
\allowdisplaybreaks[1]
\AtBeginDvi{
  
}

\renewcommand{\vec}[1]{{\bf #1}}

\newcommand\Veff{V_{\rm eff}} %
\newcommand\he[1]{#1^\dagger}%
\newcommand{\Li}[1]{\,\mbox{Li}_{#1}}
\newcommand{\Ls}[1]{\,\mbox{Ls}_{#1}}

\renewcommand{\nn}{\nonumber \\}
\renewcommand{\rmi}[1]{{\mbox{\scriptsize #1}}}
\newcommand{\rmii}[1]{{\mbox{\tiny\rm{#1}}}}

\newcommand{\CF}{C_\rmii{F}}

\newcommand{\Tc}{T_{\rm c}}

\newcommand{\Tp}{T_{\rm p}}
\newcommand{\Hp}{H_{\rm p}}
\newcommand{\alphap}{\alpha_{\rm p}}
\newcommand{\yc}{y_{\rm c}}
\newcommand{\xc}{x_{\rm c}}
\newcommand{\yn}{y_{\rm n}}
\newcommand{\yp}{y_{\rm p}}
\newcommand{\cs}{c_s}
\newcommand{\vw}{v_w}
\newcommand{\mD}{m_\rmii{D}}

\newcommand{\bmu}{\bar{\mu}}
\newcommand{\Lamd}{\bmu_{3}}
\newcommand{\LamD}{\bmu}
\newcommand{\gammaE}{{\gamma_\rmii{E}}}
\newcommand{\alephL}{\aleph_{\resizebox{6pt}{!}{\ToprVW(\Agl1,\Agl1,\Agl1,\Lgl1,\Lgl1,\Lgl1)}}}

\newcommand{\mA}{m_\rmii{$A$}}

\newcommand{\article}{{\em article}}

\newcommand{\Tint}[1]{{\hbox{$\sum$}\!\!\!\!\!\!\!\int\,}_{\!\!\!\!\raise-0.9ex\hbox{$\scriptstyle{#1}$}}}
\newcommand{\Tinti}[1]{{{\Sigma}\!\!\!\!\raise0.3ex\hbox{$\int$}_\rmii{${#1}$}}}
\newcommand{\Tintip}[1]{{{\Sigma'}\!\!\!\!\!\raise0.3ex\hbox{$\int$}_\rmii{${#1}$}}}

\def\scfc{0.7}  
\def\phgt{21}   
\def\pwc{21}    
\def\pwcb{31.5} 
\def\pwcc{42} 
\newcommand{\PIC}[4]{\;\parbox[c]{#2 pt}{\begin{picture}(#2,#3)(0,0)
      \SetWidth{1.0}\SetScale{#4} #1 \end{picture}}\;}
\renewcommand{\pic}[1]{\PIC{#1}{\pwc}{\phgt}{\scfc}}
\renewcommand{\picb}[1]{\PIC{#1}{\pwcb}{\phgt}{\scfc}}
\renewcommand{\picc}[1]{\PIC{#1}{\pwcc}{\phgt}{\scfc}}
\newcommand{\te}{\textemdash}
\newcommand{\hsq}{\Phi^\dagger \Phi}

\begin{document}

\newcommand{\HY}{\affiliation{
    Department of Physics and Helsinki Institute of Physics,
    P.O.~Box 64, FI-00014 University of Helsinki,
    Finland
}}

\newcommand{\GU}{\affiliation{
    Institute for Theoretical Physics, Goethe Universit{\"a}t Frankfurt,
    60438 Frankfurt, Germany
}}

\newcommand{\Nordita}{\affiliation{
    Nordita,
    KTH Royal Institute of Technology and Stockholm University,
    Hannes Alfv\'ens v\"ag 12,
    SE-106 91 Stockholm,
    Sweden
}}

\newcommand{\Shanghai}{\affiliation{
    Tsung-Dao Lee Institute \& School of Physics and Astronomy,
    Shanghai Jiao Tong University, Shanghai 200240, China
}}

\newcommand{\Shanghaip}{\affiliation{
    Shanghai Key Laboratory for Particle Physics and Cosmology,
    Key Laboratory for Particle Astrophysics and Cosmology (MOE),
    Shanghai Jiao Tong University, Shanghai 200240, China
}}

\newcommand{\Uppsala}{\affiliation{
    Department of Physics and Astronomy, Uppsala University,
    Box 516, SE-751 20 Uppsala,
    Sweden
}}

\title{%
  Cosmological phase transitions at three loops:
  the final verdict on perturbation theory
}

\preprint{HIP-2024-15/TH}

\author{Andreas Ekstedt\,\orcidlink{0000-0002-2240-5557}}
\email{andreas.ekstedt@physics.uu.se}
\Uppsala

\author{Philipp Schicho\,\orcidlink{0000-0001-5869-7611}}
\email{schicho@itp.uni-frankfurt.de}
\GU

\author{Tuomas V.~I.~Tenkanen\,\orcidlink{0000-0002-3087-8450}\,}
\email{tuomas.tenkanen@helsinki.fi}
\HY \Nordita \Shanghai \Shanghaip

\begin{abstract}
  \noindent
  We complete the perturbative program for
  equilibrium thermodynamics of cosmological first-order phase transitions
  by determining the finite-temperature effective potential of
  gauge-Higgs theories at
  next-to-next-to-next-to-next-to-leading order (N$^4$LO).
  The computation of
  the three-loop effective potential required to reach this order
  is extended to generic models
  in
  dimensionally reduced effective theories
  in a companion article~\cite{Ekstedt:2023xxx}.
  Our N$^4$LO result is the last perturbative order
  before confinement renders
  electroweak gauge-Higgs theories non-perturbative at four~loops.
  By contrasting our analysis with non-perturbative lattice results, we
  find a remarkable agreement.
  As a direct application for predictions of gravitational waves produced by a first-order transition, our computation provides
  the final fully perturbative results for
  the phase transition strength and
  speed of sound.
\end{abstract}

\maketitle

\section{Introduction}
Phase transitions are milestones in the early universe\te
be it by
triggering inflation~\cite{Guth:1980zm,Achucarro:2022qrl},
sparking Baryogenesis~\cite{Rubakov:1996vz,Morrissey:2012db,Bodeker:2020ghk},
or
creating a symphony of gravitational waves
\cite{Hogan:1986dsh,Caprini:2007xq,Hindmarsh:2015qta,Caprini:2019egz}.
Such transitions typically occur at high temperatures
and their properties are often difficult
to predict reliably.
Especially
for gravitational-wave (GW) production, where theoretical calculations can misjudge the peak amplitude by ten orders of magnitude~\cite{Croon:2020cgk}.
And with next-generation GW experiments\te%
LISA~\cite{Audley:2017drz},
DECi-hertz Interferometer GW Observatory (DECIGO)~\cite{Kudoh:2005as,Kawamura:2011zz},
Big Bang Observer~\cite{Yagi:2011wg},
TAIJI~\cite{Gong:2014mca},
and TIANQIN~\cite{TianQin:2015yph}%
\te
on the horizon,
it is clear that theoretical predictions are not yet up to par.
As such there is a drive to improve the predictions and augment
conventional frameworks~\cite{Quiros:1999jp,Grojean:2006bp,Delaunay:2007wb,Profumo:2007wc,Espinosa:2008kw,Kehayias:2009tn}
with state-of-the-art tools.
The tool in question is effective field theory
(EFT)~\cite{Braaten:1995cm,Braaten:1995jr,Kajantie:1995dw,Gould:2021ccf}.

These theoretical difficulties manifest when considering
GW production during a primordial phase transition.
As the universe cools down,
first-order phase transitions occur by nucleating bubbles of
the new phase within the old one.
For the electroweak case,
these bubbles release an enormous amount of latent heat and are rapidly accelerated.
As bubbles collide and generate sound waves,
a quadrupole moment is induced,
which subsequently sources gravitational waves.
The production and propagation of bubbles in a first-order phase transition is a classical process.
It occurs on length scales much larger than the temperature, {\em viz.}\ $L\gg T^{-1}$, while in the fundamental theory quantum fluctuations dominantly occur at $L\sim T^{-1}$. Conversely, making {\em classical} predictions directly within the {\em quantum} theory leads to a host of problems.
These problems include
the presence of large loop corrections and
ad-hoc recipe for calculating thermodynamic and dynamic quantities; see the discussions in~\cite{Lofgren:2023sep,Gould:2021ccf}.

A rigorous solution is to make all predictions directly within a classical theory.
To this end,
quantum fluctuations
with a characteristic energy
$E\sim T$,
are encoded in the classical theory as effective parameters. Since equilibrium dynamics is, by definition, time-independent, the effective classical theory is static and fields only depend on three-dimensional (3D) spatial coordinates. Thus the name dimensional reduction~\cite{Ginsparg:1980ef,Appelquist:1981vg}.
Thermally induced bubble nucleation is also a classical process and the rate of nucleation, $\Gamma$,
is set by the energy-cost of nucleating a bubble at rest,
$\Gamma\sim e^{-E_\rmi{bubble}/T}$.
This energy is again a static quantity, calculable within
the dimensionally reduced theory (3D~EFT).
The construction of such 3D~EFTs is a well-understood, purely perturbative process with ample applications
beyond the Standard Model (SM)~\cite{%
  Kajantie:1995kf,Cline:1996cr,Laine:1996ms,Losada:1996ju,Rajantie:1997pr,
  Andersen:1998br,Laine:2012jy,Schicho:2021gca,Niemi:2021qvp,Croon:2020cgk,
  Ekstedt:2022bff,Lewicki:2024xan,Bandyopadhyay:2021ipw,Jangid:2023jya}.

By computing
Feynman diagrams such 3D~EFTs can be studied perturbatively.
At least up to a fixed loop order.
One complication is that 3D field theories are confining
which results in glueball-like
bound states that become important at high enough loop orders,
which necessitates lattice simulations~\cite{%
  Farakos:1994xh,Kajantie:1995kf,Kajantie:1996qd,Laine:2000rm,Kainulainen:2019kyp,Gould:2019qek,Niemi:2020hto,Gould:2021dzl,Niemi:2024axp}.
For ${\rm SU}(N)$ gauge theories,
non-perturbative effects become important at four loops.
Intuitively, such non-perturbative effects are related to
the logarithmic dependence of the vector potential in
(2+1)~dimensions,%
\footnote{
  In $(d-1)+1$-dimensions,
  the potential behaves as $r^{3-d}$.
}
$V(r)\approx g^2 T\ln r$,
where $g$ is a generic coupling constant.
This gives rise to bound-states with a characteristic mass
$m_\rmii{M} \sim r_{\rm c}^{-1}\sim g^2 T$ and
the emergence of confinement at a scale
$r_{\rm c}$.
In the broken-Higgs phase this is not an issue as
$(g \phi)^{-1} \ll r_{\rm c}$
for a non-zero classical scalar background $\phi$.
In the symmetric phase, however, gauge-boson fluctuations are controlled by
the magnetic mass
$m_\rmii{M}$.
Since the free-energy has units of mass cubed in 3D,
the non-perturbative contribution is of
the order
$F^\text{non-pert}_\text{free}\sim m_\rmii{M}^3\sim g^6 T^3$
in the symmetric phase.
This is the same order as four-loop diagrams in the broken phase.
Thus non-perturbative contributions can only be ignored up to three loops.

Alternatively, in the original argument by Linde~\cite{Linde:1980ts},
such a breakdown of perturbation theory
can be seen directly by estimating the size of higher-order loops.
Since the
free-energy within the 3D theory
is computed via $\ell$-loop vacuum diagrams
linked
by $\ell-1$ vertices with 3D coupling $g^2_3 \sim g^2 T$ and
by $2\ell-2$ propagators with mass $m_\rmii{M}$
(using here $m_\rmii{M}$ for dimensional reasons),
the $\ell$-loop integration in 3D yields $m_\rmii{M}^{3\ell}$ and
renders the overall diagram proportional to $g^6_3 (g^2_3/m_\rmii{M})^{\ell-4}$.
Consequently, a magnetic-scale mass
$m_\rmii{M} \sim g^2 T$
contributes at $\mathcal{O}(g^6 T^3)$
{\em irrespective} of the loop order $\ell$.
This results in a problem deep in the infrared (IR), while modes with masses $m \gg m_\rmii{M}$ can still be treated perturbatively.

The perturbative program
aims to determine all perturbative orders before facing the IR problem.
In hot QCD, this program has a long history.
While the leading-order (LO) pressure
$p_0$ is described by the Stefan-Boltzmann law,
perturbative corrections
to the pressure $p/p_0$ were computed at
$\mathcal{O}(g^2)$~\cite{Shuryak:1977ut},
$\mathcal{O}(g^3)$~\cite{Kapusta:1979fh},
$\mathcal{O}(g^4\ln\frac{1}{g})$~\cite{Toimela:1982hv},
$\mathcal{O}(g^5)$~\cite{Arnold:1994ps,Arnold:1994eb}.
The final perturbative
$\mathcal{O}(g^6\ln\frac{1}{g})$ was
achieved already two decades ago~\cite{Kajantie:2002wa}.
At this final order,
also massless ${\rm O}(N)$ scalar field theories
were studied perturbatively~\cite{Gynther:2007bw},
and non-perturbatively using numerical methods~\cite{Hietanen:2004ew,DiRenzo:2006nh}.

This \article{} pushes the perturbative program
in electroweak theories to its limit by studying the
phase structure of
${\rm SU}(2)$ and
${\rm U}(1)$ Higgs-gauge theories at three loops.%
\footnote{
  For ${\rm SU}(N)$ + adjoint Higgs theory, which is the relevant EFT of QCD,
  similar
  perturbative computation reached four-loop level~\cite{Kajantie:2003ax};
  cf.\ also~\cite{Kajantie:1997tt,Kajantie:1998yc,Kajantie:2002pu,DiRenzo:2008en,Hietanen:2008tv}.
}
This endeavor is powered by EFT techniques combined with
the renormalization group, which allows for an all-order resummation of leading logarithms~\cite{Farakos:1994kx}.
We have also automated
the three-loop calculations for generic models,
which together with the technical details are relegated to
a companion paper~\cite{Ekstedt:2023xxx}.
Analytic, three-loop, results are provided
for scalar condensates and critical mass which
in turn can be related to the phase transition
latent heat and critical temperature.
We also compare our results with lattice Monte-Carlo simulations \cite{Gould:2022ran}, and
find that three-loop corrections significantly
improve the agreement with the lattice in
the perturbative regime.

The \article{} is organized as follows.
Section~\ref{sec:3loop:IR} defines the effective theory of interest and
describes the organization of the three-loop computation.
Section~\ref{sec:results}
presents
the results for the critical mass and scalar condensates and
compares the analytic results to previous non-perturbative lattice simulations.
In sec.~\ref{sec:GW},
we apply our computation to illuminating setups of dark sector phase transitions and discuss the impact of higher-order corrections to GW predictions.
We summarize our findings in sec.~\ref{sec:conclusions} and discuss future directions.
Appendix~\ref{sec:EFT:xpansion} organizes the broken-phase perturbative series.
Appendix~\ref{sec:u1:results} collects the thermodynamic results for the Abelian Higgs model.
Appendix~\ref{sec:EFT:su2} details the EFT construction for a simplified model.

\section{Three-loop computation}
\label{sec:3loop:IR}

The construction of dimensionally reduced effective theories
from generic parent theories is detailed
in~\cite{Kajantie:1995dw}, and
further automated in \cite{Ekstedt:2022bff}.
Here,
we directly start with the three-dimensional theory action
\begin{align}
\label{eq:S3d}
S_3=\int_{\vec{x}} \biggl[
\frac{1}{4} F_{ij}^a F_{ij}^a
+ (D_i \Phi)^\dagger (D_i \Phi)
+ V(\Phi)\biggr]
\;,
\end{align}
where
$\int_{\vec{x}} = \int {\rm d}^{d} x$ and
$d = 3 - 2\epsilon$.
The gauge coupling is denoted as $g_3$ inside the covariant derivative
$D_i\Phi = (\partial_i - i g_3 A_i)\Phi$, where
$A_i = A_i^a T^a$ and $T^a$ are the generators of ${\rm SU}(N)$
under which
the scalar
$\Phi$ transforms as an $N$-tuplet.
For the broken phase, we focus on the case
where $\Phi$ transforms
as a doublet under ${\rm SU}(2)$
and,
in appendix~\ref{sec:u1:results},
as a singlet
under ${\rm U}(1)$.
For the symmetric phase,
we retain a general $N$.
The tree-level potential is
\begin{align}
\label{eq:V:3d}
V(\Phi)=
m_{3}^{2} \hsq
+\lambda_{3}^{ } (\hsq)^2
\,,
\end{align}
with the scalar mass parameter squared $m_3^{2}$ and
scalar self-coupling $\lambda_3$.
The subscript on these couplings reminds us that these are effective parameters of the 3D~EFT, and that they have dimension of mass.
While eq.~\eqref{eq:S3d} describes the Standard Model at high temperatures and
vanishing hypercharge coupling,
many Standard Model extensions also map to this theory,
albeit with different values of couplings.
An example of such a mapping is shown in appendix~\ref{sec:EFT:su2}.

The potential in eq.~\eqref{eq:V:3d} does not exhibit a barrier.
However,
a barrier can be generated
if the vector-boson mass is large.
For this to happen,
vector-boson induced loops need to be comparable with
the tree-level potential.
Formally, the vector bosons can be integrated out in the broken phase.
In the symmetric phase,
where vector bosons are massless,
the full theory must be considered.

As a consequence,
the LO broken-phase action for ${\rm SU}(2)$ contains only
scalar fields~\cite{Ekstedt:2022zro,Gould:2023ovu}
\begin{align}
S_\rmii{LO} &=
\int_{\vec{x}}\Bigl[
  (\partial_i \Phi)^\dagger( \partial_i \Phi)
  + V_\rmii{LO}(\Phi)
\Bigr]
\,,
\nn
\label{eq:VLO}
V_\rmii{LO}(\Phi) &=
  m_{3}^2 \hsq
  + \lambda_3 (\hsq)^2
  - \frac{g_3^3}{2\pi}\Bigl(\frac{\hsq}{2}\Bigr)^{3/2}
\,.
\end{align}
We remark that this construction only works if
$m^2_3/\mA^2 \sim \lambda_3/g_3^2\ll 1$, where
$\mA \sim g_3 \phi$ is
the broken-phase vector-boson mass and
$m^2_3 \sim \lambda_3 \phi^2$.
The LO potential in eq.~\eqref{eq:VLO}
then admits a first-order transition, and
therefore serves as a good starting point for a perturbative treatment.

\begin{figure*}
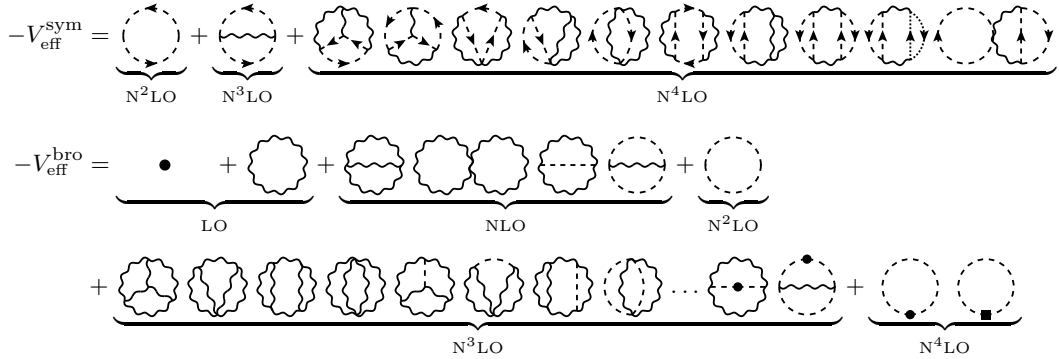

\begin{align*}
  -\Veff^{\rmi{sym}} &=
  \underbrace{
    \TopoVR(\Asc1)
  }_{\text{N$^2$LO}}
  + \underbrace{
    \ToptVS(\Asc1,\Asc1,\Lgl1)
  }_{\text{N$^3$LO}}
  + \underbrace{
    \ToprVW(\Agl1,\Asc1,\Agl1,\Lgl1,\Lsc1,\Lcs1)
    \ToprVW(\Asc1,\Agl1,\Asc1,\Lgl1,\Lcs1,\Lsc1)
    \ToprVV(\Agl1,\Asc1,\Agl1,\Lsc1,\Lcs1)
    \ToprVV(\Agl1,\Agl1,\Acs1,\Lgl1,\Lcs1)
    \ToprVB(\Agl1,\Agl1,\Acs1,\Acs1)
    \ToprVBB(\Agl1,\Acs1,\Agl1,\Acs1,\Lsc1,\Lcs1)
    \ToprVBB(\Agl1,\Agl1,\Asc1,\Agl1,\Lgl1,\Lcs1)
    \ToprVBB(\Acs1,\Agl1,\Asc1,\Agl1,\Lcs1,\Lcs1)
    \ToprVBB(\Ahg1,\Agl1,\Asc1,\Agl1,\Lhg1,\Lcs1)
    \ToprVBT(\Acs1,\Acs1,\Agl1,\Agl1,\Lcs1)
  }_{\text{N$^4$LO}}
  \, \\[3mm]
  -\Veff^{\rmi{bro}} &=
  \underbrace{
    \Vtxz
    + \TopoVR(\Agl1)
  }_{\text{LO\vphantom{$^{2}$}}}
  + \underbrace{
    \ToptVS(\Agl1,\Agl1,\Lgl1)
    \ToptVE(\Agl1,\Agl1)
    \ToptVS(\Agl1,\Agl1,\Lsr1)
    \ToptVS(\Asr1,\Asr1,\Lgl1)
  }_{\text{NLO\vphantom{$^{2}$}}}
  + \underbrace{
    \TopoVR(\Asr1)
  }_{\text{N$^2$LO}}
  \nn[2mm] &
  + \underbrace{
    \ToprVW(\Agl1,\Agl1,\Agl1,\Lgl1,\Lgl1,\Lgl1)
    \ToprVV(\Agl1,\Agl1,\Agl1,\Lgl1,\Lgl1)
    \ToprVBB(\Agl1,\Agl1,\Agl1,\Agl1,\Lgl1,\Lgl1)
    \ToprVB(\Agl1,\Agl1,\Agl1,\Agl1)
    \ToprVW(\Agl1,\Agl1,\Agl1,\Lsr1,\Lgl1,\Lgl1)
    \ToprVV(\Agl1,\Asr1,\Agl1,\Lgl1,\Lgl1)
    \ToprVBB(\Agl1,\Agl1,\Agl1,\Agl1,\Lsr1,\Lgl1)
    \ToprVB(\Agl1,\Agl1,\Asr1,\Asr1)
    \dots
    \ToptVSoa(\Agl1,\Agl1,\Lsr1)
    \ToptVSob(\Asr1,\Asr1,\Lgl1)
  }_{\text{N$^3$LO}}
  + \underbrace{
    \TopoVRo(\Asr1)
    \TopoVRb(\Asr1)
  }_{\text{N$^4$LO}}
  \,
\end{align*}
\caption{%
  Diagrams contributing to the
  symmetric- and broken-phase effective potential up to N$^4$LO.
  In the symmetric phase, wiggly lines denote ${\rm SU}(N)$ vector bosons $A_\mu^a$, directed dotted lines ghosts, and
  directed dashed lines the Higgs $N$-tuplet, $\Phi$.
  In the broken phase and for ${\rm SU}(2)$, wiggly lines denote the vectors bosons $Z,W^{\pm}$, and dashed lines scalars $h,G,G^{\pm}$.
  Blobs correspond to mass and boxes to field renormalization insertions
  of eqs.~\eqref{eq:mass:insert}--\eqref{eq:kinetic:insert}.
  The diagrams were drawn with {\tt Axodraw}~\cite{Collins:2016aya}.
}
\label{fig:diags:Veff}
\end{figure*}

In terms of a formal power counting parameter $g$
(of an underlying parent theory),
the effective potential can be expanded as~\cite{Gould:2023ovu}
\color{black}
\begin{align}
\label{eq:veff-expansion1}
\frac{\Veff}{T^3} \sim
  \underset{\text{LO}\vphantom{^2}}{
    \fbox{$\colorbox{gray!10}{$
    \frac{g^{3}}{\pi^{ }}
    \vphantom{\frac{g^{10/2}}{\pi^{7/2}}}
    $}$}}
+
  \underset{\text{NLO}\vphantom{^2}}{
    \fbox{$\colorbox{gray!10}{$
    \frac{g^{4}}{\pi^{2}}
    \vphantom{\frac{g^{10/2}}{\pi^{7/2}}}
    $}$}}
&+
  \underset{\text{N$^2$LO}}{
    \fbox{$\colorbox{gray!10}{$
    \frac{g^{9/2}}{\pi^{5/2}}
    \vphantom{\frac{g^{10/2}}{\pi^{7/2}}}
    $}$}}
  \\[1mm] &
+
  \underset{\text{N$^3$LO}}{
    \fbox{$\colorbox{gray!10}{$
    \frac{g^{5}}{\pi^{3}}
    \vphantom{\frac{g^{10/2}}{\pi^{7/2}}}
    $}$}}
+
  \underset{\text{N$^4$LO}}{
    \fbox{$\colorbox{gray!10}{$
    \frac{g^{11/2}}{\pi^{7/2}}
    \vphantom{\frac{g^{10/2}}{\pi^{7/2}}}
    $}$}}
+ \mathcal{O}(g^6)
\,.
\nonumber
\end{align}
As mentioned in the introduction,
$\mathcal{O}(g^6)$ terms require non-perturbative input.
To compute the effective potential,
we first keep $g$ as an expansion parameter
before later choosing a more appropriate expansion parameter in the 3D theory.
Concretely and as depicted in fig.~\ref{fig:diags:Veff},
the first five orders of eq.~\eqref{eq:veff-expansion1}
originate from
\begin{align*}
  \Veff^\rmii{LO}
  &\sim g^{3}
  &&
  \text{Tree-level scalar and}
  \\ &&&\text{one-loop vector}
  \,,\\[1mm]
  \Veff^\rmii{NLO}
  & \sim g^{4}
  &&
  \text{Two-loop vector}
  \,,\\[1mm]
  \Veff^\rmii{N$^2$LO}
  & \sim g^{\frac{9}{2}}
  &&
  \text{One-loop scalar}
  \,,\\[1mm]
  \Veff^\rmii{N$^3$LO}
  & \sim g^{5}
  &&
  \text{Three-loop vector and}
  \\ &&&
  \text{subleading two-loop vector-scalar}
  \,,\\[1mm]
  \Veff^\rmii{N$^4$LO}
  & \sim g^{\frac{11}{2}}
  &&
  \text{%
    Subleading one-loop scalar
  }
  \,.
\end{align*}

For the symmetric-phase potential,
multiple diagrams contribute as depicted
in fig.~\ref{fig:diags:Veff}.
While for the first two orders
at leading
and next-to-leading order (NLO)
$\Veff\bigr|_{\rmii{LO}}^{\rmi{sym}}
=\Veff\bigr|_{\rmii{NLO}}^{\rmi{sym}} \equiv 0$,
the first non-vanishing contributions arise at
next-to-next-to-leading order (N$^2$LO),
next-to-next-to-next-to-leading order (N$^3$LO), and
next-to-next-to-next-to-next-to-leading order (N$^4$LO),
{\em viz.}
\begin{align}
\label{eq:Veff:sym}
\Veff\bigr|_{\rmii{N$^2$LO}}^{\rmi{sym}} &=
  - \frac{N}{6\pi}
\bigl(m_3^2\bigr)^{\frac{3}{2}}
  \,, \\[1mm]
\Veff\bigr|_{\rmii{N$^3$LO}}^{\rmi{sym}} &=
  - \frac{g_3^{2} m_3^2}{(4\pi)^2}\frac{N\CF^{ }}{2} \Bigl[
    - 3
    + 4\ln2
    + 2\ln\frac{m_3^2}{\Lamd^2}
  \Bigr]
\,,\\[1mm]
\Veff\bigr|_{\rmii{N$^4$LO}}^{\rmi{sym}} &=
  \frac{g_3^{4}\CF^{ }}{(4\pi)^3}
\bigl(m_3^2\bigr)^{\frac{1}{2}}
\Bigl[
    \frac{81N^2+10N-87}{24}
  + \frac{N\CF}{3}\pi^2
  \nn &
  \hphantom{{}=\frac{g_3^{4}\CF^{ }}{(4\pi)^3}\bigl(m_3^2\bigr)^{\frac{1}{2}}\Bigl[}
  + \frac{12N^2-7N+27}{6}\ln2
  \nn &
  \hphantom{{}=\frac{g_3^{4}\CF^{ }}{(4\pi)^3}\bigl(m_3^2\bigr)^{\frac{1}{2}}\Bigl[}
  + \frac{4N^2 - N + 3}{4}\ln\frac{m_3^2}{\Lamd^2}
\Bigr]
\,,
\end{align}
computed in general $R_\xi$ (or Fermi) gauge;
see~\cite{Martin:2018emo} for
the generalized gauge fixing.
Here,
$\Lamd$ is the 3D renormalization scale and
$\CF = (N^2 -1)/(2N)$ is the Casimir operator in the fundamental representation.
The full ${\rm SU}(N)$ Standard-Model symmetric pressure
including adjoint (temporal) scalars is given
by~\cite{Gynther:2005av,Gynther:2005dj,Laine:2015kra}.

The two-loop $\beta$-functions
for the mass and the vacuum running,
using
$t_3 = \ln\Lamd$,
are
\begin{align}
\label{eq:beta:m}
  \partial_{t_3}^{ }
  m_{3}^{2}
  &=
  -\frac{2}{(4\pi)^2}\frac{1}{N}\Bigl[
      \frac{\CF^{ }(4N^2-N+3)}{4} g_3^{4}
    \nn &
    \hphantom{{}=-\frac{2}{(4\pi)^2}\frac{1}{N}\Bigl[}
    + 2N(N+1)\bigl(
      \CF^{ }g_3^{2}\lambda_3^{ }
    - \lambda_3^2
    \bigr)
    \Bigr]
\,,\\
\label{eq:beta:V0}
  \partial_{t_3}^{ }
  V_{0}^{ }
  &=
  -2N\CF^{ }\frac{g_{3}^{2} m_{3}^{2}}{(4\pi)^2}
\,.
\end{align}
Since the 3D~EFT is super-renormalizable \cite{Farakos:1994kx},
the mass $\beta$-function is exact,
while the vacuum $\beta$-function will also receive a four-loop contribution.
The potential $\Veff\bigr|^{\rmi{sym}}$ is
renormalization-scale independent to the computed order.
The running at
N$^2$LO, using eq.~\eqref{eq:beta:m},
is cancelled by the explicit logarithm at N$^4$LO, and
the explicit logarithm at
N$^3$LO is cancelled by the vacuum running of eq.~\eqref{eq:beta:V0}.

After splitting the scalar field into
$\Phi \to \frac{1}{\sqrt{2}}\phi\,\delta_{i,N}+ \Phi$,
the broken-phase effective potential is composed of the diagrammatic contributions given in fig.~\ref{fig:diags:Veff}.
Henceforth,
we focus on
${\rm SU}(2)$,
for which the broken-phase potential
amounts to
\begin{align}
\label{eq:Veff:su2:LO:bro}
  \Veff\bigr|_{\rmii{LO}}^{\rmi{bro}} &=
  \frac{1}{2} m_{3}^{2} \phi^2
  + \frac{1}{4} \lambda_{3}^{ } \phi^4
  - \frac{1}{16\pi} g_{3}^{3} \phi^3
  \,, \\
  \Veff\bigr|_{\rmii{NLO}}^{\rmi{bro}} &=
  \frac{g_3^{4}\phi^2}{(16\pi)^2}\frac{3}{4} \Bigl[
  11
  - 42 \ln\frac{3}{2}
  + 34\ln\frac{\Lamd}{g_{3}\phi} \Bigl]
  \,, \\[1mm]
  \Veff\bigr|_{\rmii{N$^2$LO}}^{\rmi{bro}} &=
  -\frac{1}{12\pi}\Bigl[
  \widetilde{m}_{h}^3
  + 3 \widetilde{m}_{\rmii{$G$}}^{3}
  \Bigl]
  \,, \\[1mm]
  \Veff\bigr|_{\rmii{N$^3$LO}}^{\rmi{bro}} &=
  \frac{3g_{3}^{2}}{(16\pi)^2}\Bigl[
  m_{3}^{2}\Bigl(
    5
    + 12\ln2
    + 16 \ln\frac{\Lamd}{g_{3}\phi}
  \Bigr)
  \nn &\!\!\!\!
  \hphantom{\frac{g_{3}^{2}}{(16\pi)^2}}
  + 3\lambda_{3}^{ }\phi^2\Bigl(
    3
    + 4\ln2
    + 8\ln\frac{\Lamd}{g_{3}\phi}
  \Bigr)
  \Bigr]
  \nn &
  + \frac{g_{3}^{5}\phi}{(16\pi)^3}\alephL
  \,,\\[1mm]
  \Veff\bigr|_{\rmii{N$^4$LO}}^{\rmi{bro}} &=
  - \frac{g_{3}^{ }}{(16\pi)^2}\frac{2}{\phi}\Bigl[
      11 \widetilde{m}_{h}^3
    + 48 \widetilde{m}_{\rmii{$G$}}^3
  \Bigr]
  \nn &
  +
  \frac{6 g_{3}^{4}}{(16\pi)^3}\Bigl[
  \widetilde{m}_{h}^{ }\Bigl(
    20
    + 21\ln\frac{3}{2}
    - 17\ln\frac{\Lamd}{g_{3}\phi}
  \Bigr)
  \nn&
  \hphantom{\frac{g_{3}^{4}}{(16\pi)^3}}
  + 3 \widetilde{m}_{\rmii{$G$}}^{ }\Bigl(
    3
    + 21\ln\frac{3}{2}
    - 17\ln\frac{\Lamd}{g_{3}\phi}
  \Bigr)
  \Bigr]
  .
\end{align}
The three-loop constant used above is defined as
\begin{align}
\label{eq:alephL}
\alephL &\equiv
\frac{801}{2}
+ \frac{1843}{32}\pi^2
+ 2643\ln2
- \frac{2943}{2}\ln3
- \frac{8361}{16}\ln^2\!3
\nn&
- \frac{7353}{4}\Bigl(
\ln^2\!2
- \ln2\ln3
+ \frac{1}{2}\Li{2}\Bigl(\frac{1}{4}\Bigr)
\Bigr)
- 63\Li{2}\Bigl(\frac{1}{3}\Bigr)
\nn&
- \frac{849}{2\sqrt{2}}\bigl(\Ls{2}(4\alpha)\!-\!\Ls{2}(2\alpha)\bigr)
\nn[2mm] & \approx
791.9306523639452(1)
\,,
\end{align}
where
$\alpha = \arcsin \frac{1}{3}$.
Due to the vector Mercedes diagram
($\resizebox{14pt}{!}{\ToprVW(\Agl1,\Agl1,\Agl1,\Lgl1,\Lgl1,\Lgl1)}$)~\cite{Broadhurst:1998iq},
it contains
the transcendental functions for the
polylogarithm $\Li{s}(z)$ and
log-sine integral
\begin{align}
\Ls{j}(\theta) &=
-\int_{0}^{\theta}\!{\rm d}\tau\,\ln^{j-1} \Bigl| 2\sin\frac{\tau}{2} \Bigr|
\,,
\end{align}
where $\Ls{2}(\theta)=\frac{1}{2i}(\Li{2}(e^{i\theta}) - \Li{2}(e^{-i\theta}))$.

The resummed scalar masses for Higgs and Goldstone bosons in the EFT are
\begin{align}
  \widetilde{m}_{h}^{2}(\phi) &= \partial_{\phi}^{2} V_\rmii{LO}^{ }
  \,,&
  \widetilde{m}_{\rmii{$G$}}^{2}(\phi) &= \phi^{-1}\partial_{\phi}^{ } V_\rmii{LO}^{ }
  \,,
\\
\label{eq:mass-insertion}
  {\Pi}_{h}(\phi) &= \partial_{\phi}^{2} V_\rmii{NLO}^{ }
  \,,&
  {\Pi}_{\rmii{$G$}}(\phi) &= \phi^{-1}\partial_{\phi}^{ } V_\rmii{NLO}^{ }
  \,,
\end{align}
where we treat higher-order corrections perturbatively.

The N$^3$LO and N$^4$LO potential comes from
a direct computation of three-loop diagrams in fig.~\ref{fig:diags:Veff}.
The N$^4$LO potential can also be found by using
the background-field method~\cite{Abbott:1981ke,Denner:1994xt,Henning:2014wua}, and
we have verified that the two methods agree.
The direct three-loop computation is detailed in
the companion paper~\cite{Ekstedt:2023xxx} and was conducted
in $R_\xi$ gauge~\cite{Martin:2018emo} both
using
{\tt FeynCalc}~\cite{Shtabovenko:2023idz} and
{\tt FIRE}~\cite{Smirnov:2023yhb};
as well as
using
in-house
{\tt FORM}~\cite{Ruijl:2017dtg}
software
for Feynman diagram computations
after their generation with
{\tt qgraf}~\cite{Nogueira:1991ex} and
for three-loop integration-by-parts (IBP) identities
such as in~\cite{Schroder:2005va}.

For simplicity,
we illustrate the background-field derivation.
By integrating out the vector-boson field in the scalar-field background, corrections to the mass, coupling, and kinetic terms are generated.
In turn, these corrections produce the
N$^4$LO potential.
For example,
inserting the NLO mass corrections and field renormalization in
the
one-loop scalar diagram as in fig.~\ref{fig:diags:Veff},
gives
\begin{align}
\label{eq:mass:insert}
  \TopoVRo(\Asr1)
  &= -\frac{C_{s}}{2} {\Pi}_{s}(\phi)
  \int_{\vec{p}} \frac{1}{p^2 + \widetilde{m}_{s}^{2}}
  \,,\\
\label{eq:kinetic:insert}
  \TopoVRb(\Asr1)
  &= \frac{C_{s}}{2} \bar{\Pi}_{s}(\phi) \widetilde{m}_{s}^{2}
  \int_{\vec{p}} \frac{1}{p^2 + \widetilde{m}_{s}^{2}}
  \,,
\end{align}
where a summation over $s=\{h,G\}$
is implied and
where $C_s=1$ for the Higgs and $C_s=3$ for Goldstones.
The field renormalization corrections,
$Z_s = 1 + \bar{\Pi}_s$, are
\begin{align}
\label{eq:kinetic-insertion}
\bar{\Pi}_h &= -\frac{11}{16\pi} \frac{g_3}{\phi}
\,,&
\bar{\Pi}_{\rmii{$G$}} &= -\frac{1}{\pi} \frac{g_3}{\phi}
\,.
\end{align}
This leads to a momentum-dependent vertex $-k^2 \bar{\Pi}_{s}$
as seen in the N$^4$LO contribution of eq.~\eqref{eq:kinetic:insert}.

Using the $\beta$-functions
of eqs.~\eqref{eq:beta:m} and \eqref{eq:beta:V0},
one can confirm that the effective potential, $\Veff$, is
renormalization-scale invariant to the compute order.
The running at LO cancels explicit logarithms at NLO and N$^3$LO, and the running of N$^2$LO is compensated by explicit logarithms at N$^4$LO.
Vacuum running also cancels mass-dependent scale dependence at N$^3$LO.

Contributions of $\mathcal{O}(g^6 \ln g)$ appear at the four-loop level and
have been computed in
electrostatic QCD~\cite{Kajantie:2002wa,Kajantie:2003ax}.
Such contributions can also be included without an actual four-loop computation by
utilizing renormalization-scale invariance of the effective potential,
i.e.\
$\partial_{t_3}\Veff^{ } = 0$, and
the fact that
the 3D~EFT is super-renormalizable and
its running known exactly.
For ${\rm SU}(2)$ with a doublet scalar,
this $\mathcal{O}(g^6)$ contribution (including logarithms) is
\begin{align}
\label{eq:Veff:Og6}
  \Veff&\bigr|_{\rmii{N$^5$LO}}^{\rmi{sym}} =
    \\ +&
    \frac{1}{(4\pi)^4} \Bigl[
      \mathcal{A}
    + \frac{153}{128}g^6_3 \ln \frac{m^2_3}{\Lamd^2} \Big( -1 + 4 \ln 2 + \ln \frac{m^2_3}{\Lamd^2} \Big)
  \Bigr]
  \,, \nn[2mm]
\Veff&\bigr|_{\rmii{N$^5$LO}}^{\rmi{bro}} =
    - \frac{3 \phi^2 \lambda^2_3}{(4\pi)^2} \Bigl(
        1
      + \ln \frac{\Lamd}{3 \widetilde{m}_{h}}
      + \ln \frac{\Lamd}{3 \widetilde{m}_\rmii{$G$}}
    \Bigr)
    \\ +&
    \frac{1}{(4\pi)^4} \Bigl[
      \mathcal{B}
    + \frac{153}{256}g^6_3 \ln \frac{\Lamd}{g_3 \phi} \Bigl( 5 + 12 \ln 2 + 8 \ln \frac{\Lamd}{g_3 \phi} \Bigr)
    \Bigr]
  \,.
  \nonumber
\end{align}
The coefficients
$\mathcal{A}$ and
$\mathcal{B}$
are determined through a genuine four-loop computation and
they are functions of $m^2_3$, $g_3$, $\lambda_3$, and $\phi$
but, crucially, do {\em not} involve logarithms.
The N$^5$LO broken-phase potential can also be expressed diagrammatically
and would compose of
four-loop gauge-scalar diagrams with vanishing scalar masses,
two- and three-loop gauge-scalar diagrams with scalar mass insertions, and
pure scalar two-loop sunset diagrams, in analogy to
fig.~\ref{fig:diags:Veff}.
Diagrams with mass insertions are ultraviolet (UV) finite and do not involve
a logarithmic dependence,
i.e.\ they are captured by $\mathcal{B}$.
Since perturbative contributions do not fully describe the N$^5$LO due
to non-perturbative physics (cf.~\cite{Kajantie:2000iz,Hietanen:2008tv}),
we do not include these corrections in our analysis.
In the pure IR scalar sector,
all one-loop diagrams with higher-order mass and kinetic insertions contribute
at the next order,
$\mathcal{O}(g^\frac{13}{2})$, and are hence absent in
eq.~\eqref{eq:Veff:Og6}.

The free-energy of the broken phase is obtained by formally
expanding the potential around the LO minimum
\begin{align}
F_{\rmi{bro}} &=
\Veff (\phi_{\text{min}}) =
    \Veff\bigr|_{\rmii{LO\vphantom{$^2$}}}
  + \Veff\bigr|_{\rmii{NLO\vphantom{$^2$}}}
  + \Veff\bigr|_{\rmii{N$^2$LO}}
  \nn[1mm] &
+ \Bigl(
    \Veff\bigr|_{\rmii{N$^3$LO}}
    + \phi_1^{ } \Veff'\bigr|_{\rmii{NLO\vphantom{$^2$}}}
    + \frac{1}{2} \phi_1^{2} \Veff''\bigr|_{\rmii{LO\vphantom{$^2$}}}
  \Bigr)_{\rmii{N$^3$LO}}
  \nn[1mm] &
  + \Bigl(
      \Veff\bigr|_{\rmii{N$^4$LO}}
    + \phi_2^{ } \Veff'\bigr|_{\rmii{NLO\vphantom{$^2$}}}
    \nn[1mm] &
    \hphantom{{}\Bigl(\Veff\bigr|_{\rmii{N$^4$LO}}}
    + \phi_1^{ } \Veff'\bigr|_{\rmii{N$^2$LO}}
    + \phi_1^{ } \phi_2^{ } \Veff''\bigr|_{\rmii{LO\vphantom{$^2$}}}
  \Bigr)_{\rmii{N$^4$LO}}
\,,
\end{align}
where
primes denote derivative with respect to background field.
The minimum is expanded formally as
$\phi_{\text{min}} = [\phi_0 + \phi_1 + \phi_2 + \ldots]$
and
all terms are evaluated at the LO minimum
\begin{align}
\phi_0 = \frac{3g^3_3 + \sqrt{9g^6_3 - 1024 \pi^2 \lambda_3 m^2_3}}{32\pi \lambda_3}
  \,,
\end{align}
and where
\begin{align}
\phi_1 &\equiv -\frac{
  \Veff'\bigr|_{\rmii{NLO}}}{
  \Veff''\bigr|_{\rmii{LO}\hphantom{\rmii{N}}}}
\,,&
\phi_2 &\equiv -\frac{
  \Veff'\bigr|_{\rmii{N$^2$LO}}}{
  \Veff''\bigr|_{\rmii{LO}\hphantom{\rmii{N$^2$}}}}
\,,
\end{align}
are evaluated at $\phi_0$.

As a practical step,
rescaling
the fields
$\Phi\to g_3\Phi$ and
the potential
in eq.~\eqref{eq:VLO}
as
$V_\rmii{LO}(\Phi)\to g_{3}^{-6}V_\rmii{LO}(\Phi)$,
gives
\begin{align}
V_\rmii{LO}(\Phi) \to
  y \hsq
  + x (\hsq)^2
  - \frac{1}{2\pi}\Bigl(\frac{\hsq}{2}\Bigr)^{3/2}
\,.
\end{align}
Below we also rescale
$\Lamd \to g_{3}\Lamd$.
As a result, the theory is characterized by two dimensionless couplings
\begin{align}
\label{eq:def:xy}
x &\equiv \frac{\lambda_3}{g_3^2}
\,,&
y &\equiv \frac{m_3^2}{g_3^4}
\,.
\end{align}
In the following,
the perturbative series will be organized in powers of
$x\ll 1$ since
$y\sim x^{-1}$ in the vicinity of the phase transition.
The expansion of the effective potential becomes
\begin{align}
\label{eq:Veff:xpowers}
\frac{\Veff}{g_3^6} \sim
  \underset{\text{LO}\vphantom{^2}}{
    \fbox{$\colorbox{gray!10}{$
    x^{-3}
    \vphantom{x^{-\frac{3}{2}}}
    $}$}}
+
  \underset{\text{NLO}\vphantom{^2}}{
    \fbox{$\colorbox{gray!10}{$
    x^{-2}
    \vphantom{x^{-\frac{3}{2}}}
    $}$}}
&+
  \underset{\text{N$^2$LO}}{
    \fbox{$\colorbox{gray!10}{$
    x^{-\frac{3}{2}}
    \vphantom{x^{-\frac{3}{2}}}
    $}$}}
  \\[1mm] &
+
  \underset{\text{N$^3$LO}}{
    \fbox{$\colorbox{gray!10}{$
    x^{-1}
    \vphantom{x^{-\frac{3}{2}}}
    $}$}}
+
  \underset{\text{N$^4$LO}}{
    \fbox{$\colorbox{gray!10}{$
    x^{-\frac{1}{2}}
    \vphantom{x^{-\frac{3}{2}}}
    $}$}}
+ \mathcal{O}(1)
\,,
\nonumber
\end{align}
which
reformulates eq.~\eqref{eq:veff-expansion1}
in terms of an expansion parameter {\em within} the 3D~EFT~\cite{Ekstedt:2022zro}.
This has
the advantage that the EFT can be treated independently of any parent theory.

To summarize,
at higher orders two types of contributions arise,
\begin{itemize}
  \item[(i)]
    full powers of $x$ from integrating out heavy, UV, modes including
    all vector bosons,
  \item[(ii)]
    fractional powers of $x$ from calculating
    loops with IR modes of the transitioning scalar.
\end{itemize}

To determine the phase transition critical temperature $\Tc$, or
equivalently critical mass $\yc$,
we must find the value of $y$, given $x$,
where the free energies of the two phases coincide:
\begin{align}
\Delta F(\yc(x),x) &=
  [F_\rmi{bro} - F_\rmi{sym}](\yc(x),x)
=0
\,,
\end{align}
where for a quantity $X$,
differences between
the broken and symmetric phase are henceforth denoted as
$\Delta X = X_{\rmi{bro}} - X_{\rmi{sym}}$.
The goal is to find $\yc$ order by order in $x$ which has the added advantage that observables are manifestly renormalization-scale invariant at every order.

The entropy-difference between the two phases,
$\Delta S=\frac{{\rm d}}{{\rm d}\ln T} \Delta F(\yc,x)$,
characterizes the amount of heat released by the transition, and
thus its strength.
In the effective theory all temperature dependence is encoded in
the effective couplings.
The chain-rule can be used to rewrite temperature derivatives as
$y$ and $x$ derivatives~\cite{Gould:2019qek}
\begin{align}
\label{eq:condensates}
\Delta\langle \hsq\rangle &\equiv
\frac{\partial}{\partial y} \Delta F
\,,&
\Delta\langle (\hsq)^2\rangle &\equiv
\frac{\partial}{\partial x} \Delta F
\,,
\end{align}
and express the entropy
in terms of so-called {\em condensates}~\cite{Farakos:1994xh}
\begin{align}
\label{eq:alpha:condensates}
\Delta S &=
  \Bigl(\frac{{\rm d}\,\yc}{{\rm d}\ln T}\Bigr)\Delta\langle \hsq\rangle
+ \Bigl(\frac{{\rm d}\,x}{{\rm d}\ln T} \Bigr) \Delta\langle (\hsq)^2\rangle
\,,
\end{align}
where
$\langle \hsq\rangle$ is
the quadratic scalar condensate and
$\langle (\hsq)^2\rangle$,
the quartic condensate.

\section{Results for the critical mass and scalar condensates}
\label{sec:results}

Perturbative results for
the critical mass,
and the condensates of
${\rm SU}(2)$ with a fundamental Higgs
are
known to N$^2$LO~\cite{Ekstedt:2022zro}:
\begin{align}
  \label{eq:mcRes:su2}
  \yc &=
  \frac{1}{2^3 (4\pi)^2 x^{ }}\Bigl[
  1
  - \frac{51}{2} x \ln \tilde{\mu}_3
  - (2x)^{\frac{3}{2}}
  \Bigr]
  \,,\\[1mm]
  \label{eq:phisqRes}
  \Delta\bigl\langle \hsq \bigr\rangle_\rmi{c} &=
  \frac{1}{2^3 (4\pi)^2 x^{2}}\Bigl[
  1
  + \frac{51}{2} x
  + \frac{13}{2} (2x)^{\frac{3}{2}}
  \Bigr]
  \,,\\[2mm]
  \label{eq:phisqsqRes}
  \Delta\bigl\langle \left(\hsq\right)^2 \bigr\rangle_\rmi{c} &=
  \frac{1}{2^6(4\pi)^4 x^{4}}\Bigl[
  1
  + 51 x
  + 7 (2x)^{\frac{3}{2}}
  \Bigr]
  \,,
\end{align}
where
$\tilde{\mu}_3\equiv
e^{\tfrac{11}{34}-\tfrac{42}{34}\ln\tfrac{3}{2}} (8\pi x\Lamd) \approx
0.84 (8 \pi x \Lamd)$.

After including
the three-loop corrections to the potential,
we find the critical mass and condensates at
N$^3$LO and
N$^4$LO to be
\begin{widetext}
\begin{align}
  \yc\bigr|_{\rmii{N$^3$LO}}
  &= - \frac{x^{2}}{{2^{3}(4\pi)^2 x^{ }}}
  \biggl[
  72\ln\bigl(x^\frac{7}{6}\Lamd\pi\bigr)
  - \frac{2337}{16}
  + 282\ln2
  + \frac{\alephL}{2}
  \biggr]
  \,,\\[2mm]
  \Delta\bigl\langle\hsq\bigr\rangle_\rmi{c}\bigr\vert_{\rmii{N$^3$LO}}
  &=\frac{x^{2}}{{2^{3}(4\pi)^{2} x^{2}}}
  \biggl[
  36\ln x
  - \frac{1377}{16}
  + 90\ln2
  + \frac{\alephL}{2}
  \biggr]
  \,,\\[2mm]
  \Delta\bigl\langle\bigl(\hsq\bigr)^2\bigr\rangle_\rmi{c}\bigr\vert_{\rmii{N$^3$LO}} &=
  \frac{x^{2}}{{2^{6}(4\pi)^4 x^{4}}}
  \biggl[
  120\ln x
  + 72\ln\pi
  + \frac{4017}{8}
  + 372\ln2
  + \alephL
  \biggr]
  \,,\\[2mm]
  \yc\bigr|_{\rmii{N$^4$LO}} &=
  \frac{x^{\frac{5}{2}}}{{2^{\frac{5}{2}}(4\pi)^2 x^{ }}}\Bigl[
  51\ln x
  + \frac{1025}{4}
  + 6\pi^2
  + 197\ln2
  - 126\ln3
  \Bigr]
  \,,\\[2mm]
  \Delta\bigl\langle\hsq\bigr\rangle_\rmi{c}\bigr\vert_{\rmii{N$^4$LO}} &=
  - \frac{x^{\frac{5}{2}}}{{2^{\frac{7}{2}} (4\pi)^2 x^{2}}}
  \Bigl[
  255\ln x
  + \frac{4715}{4}
  + 30\pi^2
  + 985\ln2
  - 630\ln3
  \Bigr]
  \,,\\[2mm]
  \Delta\bigl\langle\bigl(\hsq\bigr)^2\bigr\rangle_\rmi{c}\bigr\vert_{\rmii{N$^4$LO}}
  &= -\frac{x^{\frac{5}{2}}}{{2^{\frac{9}{2}} (4\pi)^4 x^{4}}}
  \Bigl[
  102\ln x
  + \frac{2671}{8}
  + 12\pi^2
  + 394\ln2
  - 252\ln3
  \Bigr]
  \,,
\end{align}

\end{widetext}
where the constant term $\alephL$ is given in eq.~\eqref{eq:alephL}.
 
Note, that the scalar condensates are renormalization-scale invariant at every order.
Furthermore, the scale dependence of $\yc$ is fully consistent with
the exact $\beta$-function given in eq.~\eqref{eq:beta:m}.
Here, in its dimensionless form, using
$\partial_{t_3}\equiv \partial_{\ln\mu_3}$,
\begin{align}
\label{eq:betay}
  \beta_y^{ } =
  \partial_{t_3}^{ } y
  &= \frac{1}{(4\pi)^2}\frac{3}{16}\left(17+48 x-64 x^2\right)
  \,.
\end{align}

The above expressions mark the {\em final} result obtainable
within perturbation theory
since four-loop calculations
are non-perturbative~\cite{Linde:1980ts,Kajantie:2002wa,Kajantie:2003ax}.

\begin{figure}[t]
\includegraphics[width=0.5\textwidth]{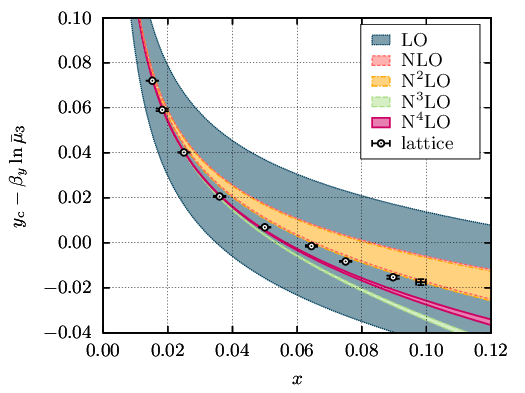}
\caption{%
  The renormalization-scale invariant quantity
  $\yc\!-\!\beta_y\ln\Lamd$, where $\yc$ is the critical mass,  $\Lamd$ is the renormalization scale, and $\beta_y$ is the $\beta$-function for
  $y$ from eq.~\eqref{eq:betay}.
  Theoretical uncertainties for lattice data from~\cite{Kajantie:1995kf,Gurtler:1997hr,Rummukainen:1998as,Laine:1998jb,Gould:2022ran} are shown as error bars,
  uncertainties for the perturbative calculations are shown as error bands by varying
  $\Lamd/(2\mA) = [10^{-\frac{1}{2}},10^{\frac{1}{2}}]$.
  The line of first-order phase transitions ends in
  a second-order transition at $x\sim 0.1$,
  indicated by the rightmost lattice point.
  The orders NLO and N$^2$LO are almost identical.
}
\label{fig:yc}
\end{figure}
\begin{figure}[t]
  \centering
  \includegraphics[width=0.5\textwidth]{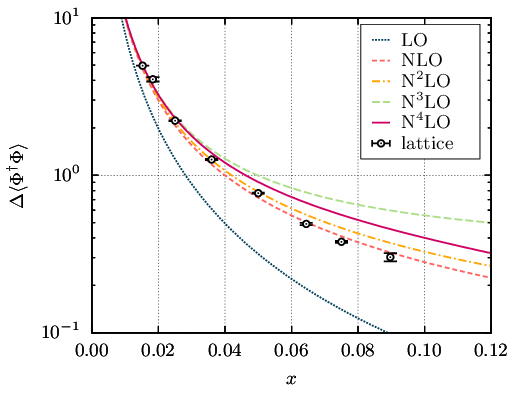}\\
  \includegraphics[width=0.5\textwidth]{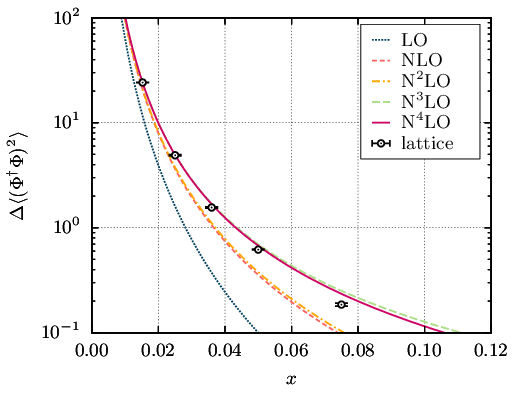}
  \caption{%
    The jumps in the
    quadratic $\Delta \langle (\Phi^\dagger \Phi)^{ } \rangle$ and
    quartic $\Delta \langle (\Phi^\dagger \Phi)^{2} \rangle$ scalar condensates as functions of $x$,
    computed in the $x$-expansion and on the lattice.
    Both condensates are manifestly gauge and renormalization-scale invariant.
    The lattice data is taken from~\cite{%
      Kajantie:1995kf,Gurtler:1997hr,Rummukainen:1998as,Laine:1998jb,Gould:2022ran}.
  }
  \label{fig:condensates}
\end{figure}
For the critical mass and the two condensates,
we contrast our perturbative results with lattice simulations~\cite{Kajantie:1995kf,Gurtler:1997hr,Rummukainen:1998as,Laine:1998jb,Gould:2022ran}
in figs.~\ref{fig:yc} and \ref{fig:condensates}.
All three quantities display a marked improvement at small $x$; this is especially poignant for $\Delta\bigl\langle(\hsq)^2\bigr\rangle$.
As a consequence,
the departure of perturbation theory from lattice data is delayed until
$x\sim 0.04-0.05$.
A subsequent breakdown of perturbation theory is indeed expected
as a second-order transition takes place at
$x = x_* = 0.0983(13)$~\cite{Gurtler:1997hr,Rummukainen:1998as}.

To illustrate the perturbative uncertainty at each order,
we varied the renormalization scale in fig.~\ref{fig:yc},
in the range
$\Lamd/(2\mA) = [10^{-\frac{1}{2}},10^{\frac{1}{2}}]$.
An optimized value for $\Lamd$ can be found \`a la
principle of
{\em minimal sensitivity}~\cite{Stevenson:1981vj,Ghisoiu:2015uza,Gould:2021dzl}.

\section{Impact on gravitational waves}
\label{sec:GW}

To investigate the importance of three-loop corrections,
we apply the results of sec.~\ref{sec:results} to a four-dimensional
parent theory that maps into eq.~\eqref{eq:S3d} in the high-temperature limit.
For maximal freedom of the analysis,
we treat our setup as a toy model for gravitational waves from a purely dark sector~\cite{Croon:2018erz}, and do not assume the Standard Model field content.
To this end, we assume a ${\rm SU}(2)$ gauge theory with a scalar doublet $\Phi$ and an additional scalar singlet $S$.
The model definition and its mapping into the thermal EFT are detailed in appendix~\ref{sec:EFT:su2}.

\hphantom{}
\paragraph*{Pressure {\rm ($p$)}.}

To derive the relevant equilibrium thermodynamic quantities for GW production, we construct the pressure in the high-temperature expansion. It is composed as
\begin{align}
\label{eq:pressure}
  p = p_0 - T \Veff
  \,,
\end{align}
where $\Veff$ is the effective potential of the 3D~EFT computed in
sec.~\ref{sec:3loop:IR}, separately for different phases.
The unit operator $p_0$, or the field-independent pressure~\cite{Braaten:1995cm},
comes from thermal corrections to the vacuum.
From the pressure,
we can find the enthalpy $\omega \equiv T p'$, the speed of sound $\cs^2 \equiv p'/e'$, and the energy density $e = T p'-p$, where we use the shorthand $p'\equiv {\rm d} p/{\rm d}T$.

To keep the discussion simple, we regroup the different perturbative orders of the pressure as
``1-loop'',
``2-loop'', and
``3-loop''.
A detailed composition of different perturbative orders of the pressure is given in appendix~\ref{sec:pert:exp}.

We consider two scenarios for the dark-sector model:
\begin{itemize}
  \item[(A)]
    If the singlet is weakly coupled to a light doublet,  the doublet can undergo a first-order phase transition while the singlet decouples.
  \item[(B)]
    Even if the doublet is too heavy to accommodate a phase transition,  a transition can be catalyzed if the interaction with the singlet is sufficiently strong.
    Here, we will work in a mass regime that admits integrating out the singlet such that only the doublet remains in the EFT.
\end{itemize}

By focusing on option~(A), we fix the parameters of the parent theory (defined in appendix~\ref{sec:4D:su2+doublet+singlet}) to
\begin{align}
\label{eq:BMA}
  g^2 &= 0.5
  \,,&
  \lambda_\phi &= 0.0075
  \,,&
  M_\phi &= 10 \sqrt{2}
  \,,
  \tag{BM-A}
\end{align}
(in arbitrary units of mass) and assume a simple tree-level relation
$\mu^2_\phi = -M^2_\phi/2$
between the doublet mass parameter and the pole mass.
The ratio $\lambda_\phi/g^2 = 0.015$ is the leading, temperature-independent contribution to the dimensionless variable $x$ that controls the behaviour of the EFT.

\begin{figure}[t]
  \centering
  \includegraphics[width=0.5\textwidth]{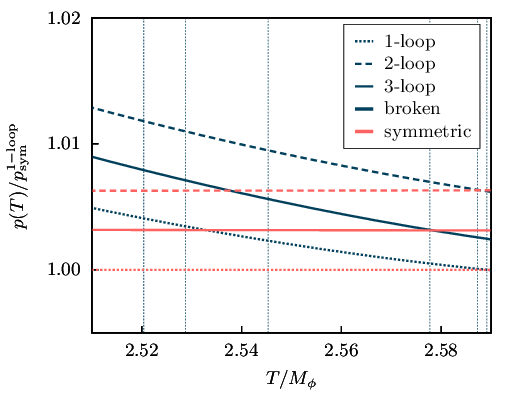}
  \includegraphics[width=0.5\textwidth]{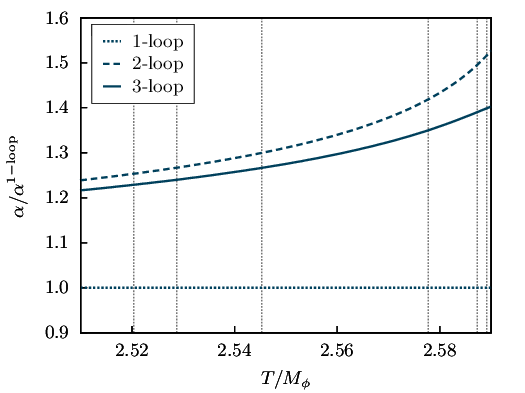}
  \caption{%
    Pressure (top) and
    phase-transition strength (bottom) as function of temperature $T$ (scaled by scalar mass $M_\phi$) in a benchmark point~\eqref{eq:BMA}.
    From left to right, vertical lines indicate the
    nucleation temperatures
    and
    critical temperatures
    of tab.~\ref{tab:TcTp:BMA}.
  }
  \label{fig:pressure}
\end{figure}%
We illustrate the pressure as a function of temperature in fig.~\ref{fig:pressure}~(top) and normalize the pressure in both phases by the one-loop symmetric-phase pressure.
Two-loop corrections affect the pressure
at the 1\% level, and subsequent three-loop corrections affect the pressure at
at the 0.1\% level, indicating excellent convergence.

\hphantom{}
\paragraph*{Critical temperature {\rm ($\Tc$)}.}

\begin{table}[t]
\begin{tabular}{|l|c|}
  \hline
  \multicolumn{2}{|c|}{$\Tc$}
  \\ \hline \hline
  1-loop &
  2.589
  \\
  2-loop &
  2.587
  \\
  3-loop &
  2.578
  \\ \hline
\end{tabular}
\hspace{1mm}
\begin{tabular}{|l|c|}
  \hline
  \multicolumn{2}{|c|}{$\Tp$}
  \\ \hline \hline
  LO
  &
  2.545
  \\
  NLO
  &
  2.529
  \\
  N$^2$LO
  &
  2.520
  \\ \hline
\end{tabular}
  \caption{%
    The critical temperature $\Tc$ and
    percolation temperature $\Tp$
    in units of the scalar mass $M_\phi$
    for
    benchmark point~\eqref{eq:BMA}
    at different perturbative orders.
  }
\label{tab:TcTp:BMA}
\end{table}

The critical temperature $\Tc$,
at each order, corresponds to
the temperatures where
the pressure-difference of the phases vanishes.
For~\eqref{eq:BMA} the critical temperatures are
listed in tab.~\ref{tab:TcTp:BMA} and
depicted by the three rightmost vertical lines
of fig~\ref{fig:pressure}.
Since the broken-phase pressure has the symmetric-phase pressure subtracted, the pressure vanishes exactly at the critical mass $\yc(\xc)$ at each order.%
\footnote{%
\label{ft:mixed-method}%
  For determining $\Tc$, we employ the {\em mixed method} of~\cite{Gould:2023ovu}. Therein, only
  the effective potential is expanded in strict perturbation theory. For $\Tc$, we do {\em not} perform a strict expansion directly
  since a similar fully strict expansion for the phase transition strength, $\alpha$, would become a functionally formidable task.
}

In the temperature window of fig.~\ref{fig:pressure},
the critical $\xc \approx 0.021$, as
contributions of temporal scalars slightly increase the value of $x$ from its
leading value set by $\lambda_\phi/g^2 = 0.015$;
see appendix~\ref{sec:EFT:su2}.
Since also $T^2/M^2_\phi \sim 6$,
the high-temperature expansion is applicable.

We have verified that points with smaller (larger)
$\lambda_\phi/g^2$ lead to increased (decreased) convergence in perturbation theory. Benchmark point~\eqref{eq:BMA} illustrates such generic trends.

\hphantom{}
\paragraph*{Phase transition strength {\rm ($\alpha$)}.}

The convergence of the pressure is inherited by all quantities further derived from it. One such quantity is the phase
transition strength defined
via~\cite{Giese:2020rtr,Giese:2020znk}
\begin{align}
  \label{eq:alpha}
  \alpha &= \frac{\Delta \bar\theta}{3\omega_\rmi{sym}}
  = \frac{1}{3 p_\rmi{sym}'} \Big(
  \frac{\Delta \Veff}{c^2_{s,\rmi{bro}}}
  - \Delta \frac{{\rm d} \Veff}{{\rm d}\ln T}
  \Big)
  \,,
\end{align}
where the pseudotrace anomaly is
$\bar\theta \equiv e - p/c^2_{s,\rmi{bro}}$ and where we used the notation
$\Delta p\equiv p\bigl|_\text{bro}-p\bigl|_\text{sym} =- T \Delta \Veff$.

In fig.~\ref{fig:pressure}~(bottom), we depict the relative-to-leading-order convergence of $\alpha$, in analogy to fig.~\ref{fig:pressure}~(top). For the pressure, the relative difference between different loop orders are minute due to the dominating unit operator $p_0$. Since $\alpha$ depends on the pressure {\em difference}, it is independent of $p_0$. Conversely, for $\alpha$, relative differences between one- and two-loop level can reach
10\%, while
the difference between two- and three-loop level
displays great convergence.

\hphantom{}

\paragraph*{Speed of sound {\rm ($\cs$)}.}

\begin{figure}[t]
  \centering
  \includegraphics[width=0.5\textwidth]{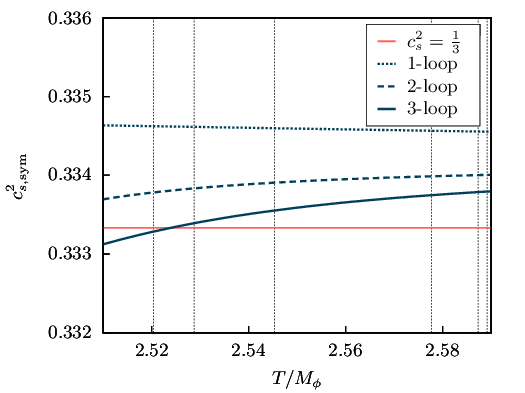}
  \includegraphics[width=0.5\textwidth]{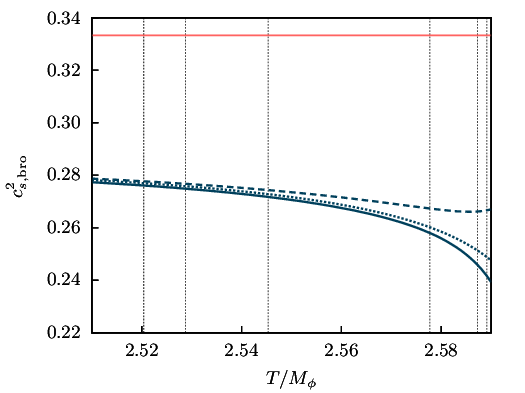}
  \caption{%
    As fig.~\ref{fig:pressure} but depicting
    the speed of sound squared, $\cs^2$,
    as a function of temperature for the
    symmetric (top) and
    broken (bottom) phase.
  }
  \label{fig:sos}
\end{figure}

While $\alpha$ is dominated by the second, derivative, term in eq.~\eqref{eq:alpha}, the 
term proportional to $\Delta \Veff$ is subdominant.
This term also depends on the broken-phase speed of sound which is plotted in fig.~\ref{fig:sos}.
The symmetric-phase result is very close to the LO result $\cs^2 = 1/3$ at all orders. The broken-phase speed of sound significantly deviates from this LO value but quickly converges with the loop order. Such deviations can be important for analyses of hydrodynamic properties of the phase transition as they could affect the shape of the GW signal and suppress it by an order of magnitude~\cite{Giese:2020znk,Giese:2020rtr}. A careful computation of the speed of sound is required especially for theories whose particle content deviates strongly from the Standard Model one~\cite{Tenkanen:2022tly}. The \article{} at hand completes the perturbative determination of the speed of sound.

\hphantom{}
\paragraph*{Bubble nucleation rate {\rm ($\Gamma$)}.}

Both figs.~\ref{fig:pressure} and ~\ref{fig:sos} depict
the results for the bubble nucleation temperature,
given by the three leftmost vertical lines
from tab.~\ref{tab:TcTp:BMA}.
These results are obtained by directly following~\cite{Ekstedt:2021kyx,Ekstedt:2022ceo};
see also~\cite{%
  Moore:2000jw,Gould:2021ccf,Lofgren:2021ogg,Hirvonen:2021zej,
  Ekstedt:2022tqk,Ekstedt:2023sqc}.
Here, we merely summarize the rationale for completeness. The bubble nucleation rate can be approximated by
$\Gamma \simeq \frac{\kappa}{2\pi} \Sigma$,
where the dynamical prefactor $\kappa$ depends
on dissipative processes~\cite{Langer:1969bc}.
Its statistical part $\Sigma$ can be computed within the 3D~EFT as
\begin{align}
\Sigma \approx [\mbox{det}_S] e^{-S_B(x,y)}
  \,,
\end{align}
where $S_B$ is the bounce action and higher-order corrections are described by the determinant $[\mbox{det}_S]$.
Henceforth, we simply work in approximations that ignore the dynamical prefactor $\kappa$~\cite{Gould:2024chm}.

In the 3D~EFT, $\Sigma$ depends only on $x,y$ and we can formally write the statistical part of the rate as an exponential
$\exp(-S_B + \ln [\mbox{det}_S]) \equiv \exp(-S_\text{eff})$, and further find the effective action in a strict expansion
\begin{align}
S_\rmi{eff} =
      S_\rmii{LO}
    + x S_\rmii{NLO}
    + x^{\frac{3}{2}} S_\rmii{NNLO}
    + \ldots
    \,.
\end{align}
Here, the powers of $x$ merely indicate the suppression relative to LO.
Expressions for each order can be found in~\cite{Ekstedt:2022ceo}.
In analogy with sec.~\ref{sec:3loop:IR},
integer powers of $x$ come from integrating out vector bosons (in the UV), while fractional powers come from scalar fluctuations (in the IR).

To relate the statistical $S_\rmi{eff}$ to cosmology and the temperature evolution of the universe, we can approximate the condition for successful percolation after bubble nucleation as
\begin{equation}
\label{eq:nucl:condition}
  S_\rmi{eff}(x,\yp) \simeq F[H(\Tp)]
  \,,
\end{equation}
where $F$ is a function of
the Hubble parameter $H(T)$.
The exact form of $F$ can be found in~\cite{Enqvist:1991xw,Gould:2022ran,Ellis:2018mja},
but here we approximate it
as a constant
$F[H(\Tp)] \approx 126$,
which corresponds to about two-thirds of the universe being in the broken phase~\cite{Ellis:2018mja}.
We remark that several definitions of $\Tp$ appear in
the literature~\cite{Caprini:2019egz,Guo:2021qcq,Croon:2020cgk}, and
subtleties concerning this reference temperature for GW production
were recently discussed in~\cite{Athron:2022mmm}.

Given $S_\rmi{eff}(x,\yp) \equiv 126$,
we can (numerically) find $\yp(x)$ as a curve in the $(x,y)$-plane.
For any parameter point of a parent theory,
the percolation temperature $\Tp$ is given by the condition
$y(x(\Tp)) = \yp(x)$.%
\footnote{
  As~\cite{Gould:2022ran},
  we employ
  $\yp$ which
  is the ``nucleation mass'' $\yn$ of~\cite{Ekstedt:2022ceo}.
}
This is analogous to finding the critical temperature from the condition
$y(x(\Tc)) = \yc(x)$~\cite{Kajantie:1995kf}.

In perturbation theory, we can find the LO result from
$y(x(\Tp^\rmii{LO})) = \yp^{\rmii{LO}}(x)$ where
$\yp^{\rmii{LO}}$ is determined by $S_\rmii{LO}$.
The higher-order corrections
$\Tp^\rmii{NLO}$ and
$\Tp^\rmii{N$^2$LO}$ are then found in a strict perturbative expansion,
as detailed in~\cite{Ekstedt:2022ceo}.
In figs.~\ref{fig:pressure} and~\ref{fig:sos},
we observe that in~\eqref{eq:BMA}, the percolation and critical temperatures are rather close. This is a generic feature of this EFT where percolation occurs with relatively little supercooling below the critical temperature.

While for the critical temperature,
we found all results including corrections up to and including N$^4$LO,
the strict expansion for the percolation temperature is performed only for
the first three orders.
The two remaining orders for the bubble nucleation rate are promoted to future work.

\hphantom{}
\paragraph*{Inverse duration {\rm ($\beta/H$)}.}

Another important thermal parameter for GW production is
the inverse duration of the transition, defined as
\begin{align}
  \frac{\beta}{H} &= - \frac{{\rm d}\ln\Gamma}{{\rm d}\ln T}
  \,.
\end{align}
For the benchmark point~\eqref{eq:BMA} of figs.~\ref{fig:pressure} and~\ref{fig:sos}, we report the values listed in tab.~\ref{tab:alpha:beta} for
the phase transition strength, $\alpha$, and
the inverse duration $\frac{\beta}{H}$
at $T=\Tp$.%
\footnote{
  For $\frac{\beta}{\Hp}$,
  we determine the first three orders in strict perturbative expansion as described in~\cite{Ekstedt:2022ceo}.
}
\begin{table}[t]
\begin{tabular}{|l|r|c|l|r|}
  \hline
  $\alpha^{\rmii{1-loop}}\bigl(\Tp^\rmii{LO}\bigr)$ &
  \vphantom{$\left(\frac{\beta}{\Hp}\right)_{\rmii{LO$_{\vphantom{x}}$}}$}
    0.014
  \\
  $\alpha^{\rmii{2-loop}}\bigl(\Tp^\rmii{NLO}\bigr)$ &
  \vphantom{$\left(\frac{\beta}{\Hp}\right)_{\rmii{NLO$_{\vphantom{x}}$}}$}
    0.020
  \\
  $\alpha^{\rmii{3-loop}}\bigl(\Tp^\rmii{N$^2$LO}\bigr)$ &
  \vphantom{$\left(\frac{\beta}{\Hp}\right)_{\rmii{N$^2$LO$_{\vphantom{x}}$}}$}
    0.021

  \\ \hline
\end{tabular}
\hspace{1mm}
\begin{tabular}{|l|r|c|l|r|}
  \hline
  $\left(\frac{\beta}{\Hp}\right)_{\rmii{LO$_{\vphantom{x}}$}}$ &
    13578
  \\
  $\left(\frac{\beta}{\Hp}\right)_{\rmii{NLO$_{\vphantom{x}}$}}$ &
    7182
  \\
  $\left(\frac{\beta}{\Hp}\right)_{\rmii{N$^2$LO$_{\vphantom{x}}$}}$ &
    4329
  \\ \hline
\end{tabular}
  \caption{%
   The phase transition strength, $\alpha$, and
   inverse duration $\frac{\beta}{\Hp}$ for
   the benchmark point~\eqref{eq:BMA} of
   figs.~\ref{fig:pressure} and ~\ref{fig:sos}.
   Percolation temperatures correspond to tab.~\ref{tab:TcTp:BMA}.
    }
\label{tab:alpha:beta}
\end{table}

Higher-order corrections increase $\alpha$ and the overall convergence is good, while they decrease $\frac{\beta}{\Hp}$ and convergence is less pronounced, even for the relatively small value of $x\approx 0.021$ of~\eqref{eq:BMA}.
The underlying perturbative computation of the bubble nucleation rate
breaks down for $x > 0.058$. We have verified improved convergence for parameter points that map into smaller values of $x$.
Furthermore, we have verified that the aforementioned trends for $\alpha$ and $\frac{\beta}{\Hp}$ also hold for other parameter points.

\hphantom{}
\paragraph*{Gravitational waves.}

\begin{figure}[t]
  \centering
  \includegraphics[width=0.5\textwidth]{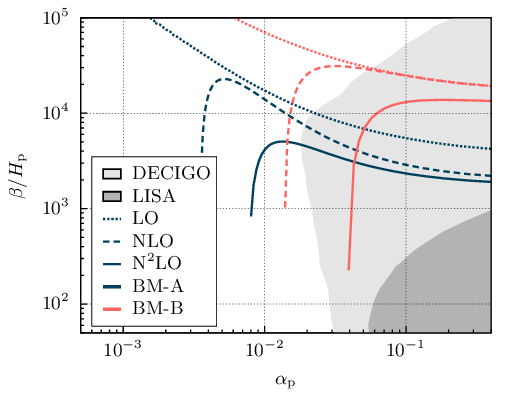}
  \caption{%
  Connecting our thermal computation to experimental probes via the
  $(\alphap,\beta/\Hp)$-plane by projecting results for
  benchmark points~\eqref{eq:BMA} but by varying
  $\lambda_\phi \in [0.0001, 0.03]$ ($x \in [0.0054,0.068]$) and
  \eqref{eq:BMB} by varying $\lambda_m \in [2,2.55]$ ($x \in [0.017,0.098]$).
  Also shown are the tentative
  sensitivity regions for
  LISA~\cite{Caprini:2019egz} and
  DECIGO~\cite{Kawamura:2011zz}.
  For these curves,
  $x$ decreases from left to right.
  For large $x$,
  the perturbative expansion for the bubble nucleation rate breaks down,
  resulting in tails of rapidly falling $\beta/\Hp$.
  }
  \label{fig:alpha-beta}
\end{figure}

After detailing several phase transition thermodynamic quantities over a range of temperatures,
we now emulate a scan over the model parameter space,
typically performed in studies of cosmological phase transitions in
beyond the Standard Model
theories or dark sectors;
cf.~\cite{Caprini:2019egz} and references therein.
Along the lines of~\cite{Caprini:2019egz},
we recast our scan results in
the $(\alpha,\beta/H)$-plane in fig.~\ref{fig:alpha-beta} for
the benchmark point~\eqref{eq:BMA},
varying
$\lambda_\phi \in [0.001, 0.03]$ which corresponds to
$x \in [0.006,0.068]$ within the EFT.
For illustration,
we show tentative sensitivity regions of LISA and DECIGO in
analogy with~\cite{Friedrich:2022cak}.%
\footnote{%
  For connecting with LISA-generation experimental probes~\cite{LISACosmologyWorkingGroup:2022jok}, we henceforth assume the scalar mass $M_\phi$ and all other dimensionful quantities in units of GeV.
  The tentative integrated sensitivity regions
  at a wall velocity $\vw=0.95$ for
  LISA at ${\rm SNR} = 5$~\cite{LISA:2017pwj} with $\mathcal{T}=4$~year mission duration~\cite{Caprini:2019egz} and
  DECIGO with the {\em Correlation} design~\cite{Kawamura:2011zz} were
  taken from~\cite{Friedrich:2022cak}.
}
While none of the transitions have
sufficiently large timescales
for LISA to probe them,%
\footnote{
  This result is expected since phase transitions that map into he 3D~EFT of ${\rm SU}(2)$ with a doublet are not observable by LISA~\cite{Gould:2019qek}.
}
DECIGO and other future GW observatories could observe a primordial GW echo from a dark sector phase transition such as those analyzed here.

Parameter points with small $\lambda_\phi$ result in small $x$, which provide the
largest (smallest) value for $\alphap$ ($\beta/\Hp$) towards the bottom right. Even for the smallest values of $x$, perturbative corrections to $\beta/\Hp$ are sizeable. When $x$ increases, the perturbative computation for the inverse duration breaks completely, as indicated by
the deviation of different orders towards
the top left.
Encouragingly, perturbation theory behaves best towards and within the DECIGO sensitivity, and a total breakdown of perturbation theory occurs only for weaker transitions, which is to be expected.

In fig.~\ref{fig:alpha-beta} and in addition to the benchmark point~\eqref{eq:BMA}~(blue), we also report on another benchmark point~\eqref{eq:BMB}~(red),
{\em viz.}
\begin{align}
\label{eq:BMB}
  g^2 &= 0.9
  \,,&
  \lambda_s &= 1.0
  \,,&\!
  \sin \theta &= 0.2
  \,,
  \nn
  \frac{M_\phi}{\rm GeV} &= 125
  \,,&
  \frac{M_s}{\rm GeV} &= 325
  \,,&\!
  \frac{v_0}{\rm GeV} &= 246
  \,,
  \tag{BM-B}
\end{align}
where parameters are defined in appendix~\ref{sec:4D:su2+doublet+singlet}.
This scenario emulates the Standard Model
augmented with a new scalar field.
By design, the singlet is sufficiently heavy to be integrated out from the final EFT; see appendix~\ref{sec:EFT:su2}.
This serves as an illuminating example that
the 3D~EFT of ${\rm SU}(2)$ with a doublet is an invaluable tool for studying the thermodynamics of electroweak-like first-order phase transitions in
the presence of field content beyond
the dynamical degrees of freedom of
the final EFT~\cite{Cline:1996cr,Laine:2012jy,Brauner:2016fla,Andersen:2017ika,Niemi:2018asa,Gorda:2018hvi,Gould:2019qek}.

\section{Discussion}
\label{sec:conclusions}

This study delivers the final verdict of perturbation theory on electroweak gauge-Higgs theories at high temperature. By accounting for three-loop contributions, any further study perforce requires non-perturbative analyses due to the Linde problem. This \article{} builds upon the recent formulation of the re-reorganized perturbative expansion~\cite{Ekstedt:2022zro}. While~\cite{Ekstedt:2022zro} computed its first three orders, here we deliver the remaining two perturbative orders and herald the culmination of a three-decade program initiated by~\cite{Arnold:1992rz,Farakos:1994kx} for the perturbative analysis of cosmological and electroweak phase transitions in particular.

A similar perturbative completion was previously only achieved for
pure scalar theory
and QCD at high temperatures.
For both theories,
perturbative results
are in remarkable agreement with non-perturbative
lattice Monte-Carlo analyses~\cite{Gould:2021dzl,Kajantie:2003ax}.
In turn, for electroweak gauge-Higgs theories,
we find remarkable agreement between perturbation theory and lattice simulations
by reorganizing the perturbative expansion.
This provides additional evidence that
the Linde problem is of minor significance in
cosmological phase transitions driven by a fundamental scalar field.

Naturally,
the precision of our perturbative results diminishes
as we approach the critical point.
At this juncture,
the transition becomes second-order and subsequently
evolves into a crossover.
Perturbation theory is inherently unreliable for such weak transitions.
One estimate suggests that perturbation theory can be trusted
for positive critical masses, $\yc$.
However, once $\yc$ turns negative,
a second-order transition is expected within perturbation theory.

The perturbative toolbox completed in this \article{} readily generalizes to
more complicated theories with multiple scalar fields \cite{Gould:2023ovu}.
Thus,
computing large perturbative higher-order corrections can
follow the methodology outlined in this \article{}.
In particular,
by utilizing the general mass hierarchy between
the broken-phase vector bosons and
the light transitioning scalar(s),
we reorganized the theory to all perturbative orders.

For predictions of
a stochastic GW background from a dark sector phase transition,
we found that higher-order perturbative corrections ensure a well-controlled perturbation theory for equilibrium thermodynamics. However, the remaining significant uncertainty stems from the current inability to determine the bubble nucleation rate with similar precision.
The extension of our findings to more complicated and
widely studied
theories with non-minimal scalar sectors, and their implications for
collider phenomenology and GW predictions, is left to future studies.

\hphantom{}
\paragraph*{Outlook.}
With the perturbative program now completed,
there is ample opportunity for future lattice studies.
While perturbation theory can reliably predict the properties of strong transitions,
it falls short in describing weak transitions where non-perturbative
effects are large.
To reliably rule out the possibility of a first-order transition,
non-perturbative lattice analyses are essential.
These analyses are particularly important for the phenomenological study of
extensions of the Standard Model~\cite{Niemi:2024axp} and
understanding their potential signatures at future colliders~\cite{Ramsey-Musolf:2019lsf}.
To this end, not only
deciding the phase-transition character with lattice techniques is important,
also future simulations for
bubble nucleation~\cite{Moore:2000jw,Moore:2001vf,Gould:2022ran,Gould:2024chm} and Sphaleron rates~\cite{Moore:1998swa,DOnofrio:2014rug,Annala:2023jvr}
are required.

To further enhance precision
in understanding cosmological phase transitions,
we consider the following future directions:
\begin{itemize}
  \item[\resizebox{17pt}{!}{$\Vtxs(\Lsr1,\Lsr1,\Lsr1,\Lsr1,\Lsr1,\Lsr1)$}]
    At $\mathcal{O}(g^6)$,
    additional contributions arise
    in the dimensional reduction via higher-dimensional operators within
    the 3D~EFT.
    While evaluated for the pure gauge sector of QCD~\cite{Laine:2018lgj},
    accurately determining their effect in generic theories remains important for
    probing the validity of the 3D~EFT.
    This validity forms a backbone not only for perturbative, but also lattice studies.
  \item[\resizebox{17pt}{!}{$\TopfVH(\Asai,\Asai,\Asai,\Asai,\Lsai,\Lsai,\Lsai,\Lsai,\Lsai)$}]
    Purely perturbative $\mathcal{O}(g^6)$ contributions from the hard scale could still be computed for the symmetric-phase pressure;
    cf.\ progress in
    hot QCD~\cite{Navarrete:2022adz}.
  \item[\resizebox{17pt}{!}{$\ToptVTpT(\Asa1,\Asa1,\Lsa1)$}]
    Building on~\cite{Ekstedt:2022ceo,Gould:2021oba},
    the final perturbative corrections for
    the thermal bubble nucleation rate and
    the phase transition duration are yet to be computed.
    They still source one of
    the largest uncertainties for
    predicting GW signals from cosmological phase transitions.
\end{itemize}

\begin{acknowledgments}
  We are grateful to
  Oliver Gould,
  Joonas Hirvonen,
  Maciej Kierkla,
  Johan L\"ofgren,
  Lauri Niemi,
  Daniel Schmitt,
  Bogumi{\l}a {\'S}wie{\.z}ewska,
  Van Que Tran,
  Jorinde van de Vis,
  Yanda Wu
  and Guotao Xia
  for discussions as this project was carried out.
  The work of AE has been supported by the Swedish Research Council, project number VR:$2021$-$00363$.
  PS acknowledges support by
  the Deutsche Forschungsgemeinschaft (DFG, German Research Foundation) through
  the CRC-TR 211
  `Strong-interaction matter under extreme conditions' --
  project number 315477589 --
  TRR 211.
  PS acknowledges the hospitality of CERN during the final stages of this work.
  TT was supported under National Natural Science Foundation of China grant 12375094.
\end{acknowledgments}

\appendix
\allowdisplaybreaks

\section{Expansion of broken phase effective potential in terms of EFT matching}
\label{sec:EFT:xpansion}

This appendix illustrates the organization of
the perturbative series of fig.~\ref{fig:diags:Veff}
for the broken-phase effective potential,
in terms of an EFT matching between
the mass scale of the vector bosons and
the scale of the phase transition.
By adopting the terminology of~\cite{Gould:2023ovu},
these scales are dubbed
{\em soft} and
{\em supersoft}.

The supersoft scale EFT is constructed by integrating out the soft scale.
The resulting action is
\begin{align}
  \label{eq:ss:tree}
  S_{\rmi{supersoft}}^{\rmi{tree}} &=
  \int_{\vec{x}} \frac{1}{2} (\partial_i s)^2 Z_s
  + \underbrace{
  	\Vtxz
  	+ \TopoVR(\Agl1)
  }_{\text{LO}}
  \nn[2mm] &
  + \underbrace{
  	\ToptVS(\Agl1,\Agl1,\Lgl1)
  	\ToptVE(\Agl1,\Agl1)
  	\ToptVS(\Agl1,\Agl1,\Lsr1)
  	\ToptVS(\Asr1,\Asr1,\Lgl1)
  }_{\text{NLO}}
  \nn[2mm] &
  + \underbrace{
  	\ToprVW(\Agl1,\Agl1,\Agl1,\Lgl1,\Lgl1,\Lgl1)
  	\dots
  	\ToptVSoa(\Agl1,\Agl1,\Lsr1)
  	\ToptVSob(\Asr1,\Asr1,\Lgl1)
  }_{\text{N$^3$LO}}
  \,,
\end{align}
where
the summation over the $s$ soft scalar fields is implied.
The construction of the EFT action follows a standard EFT
matching computation using
the background-field method~\cite{Abbott:1981ke,Denner:1994xt,Henning:2014wua};
see also~\cite{Hirvonen:2022jba} for an intuitive explanation.
In this framework,
the strategy is to
integrate only over soft, UV, momenta and
expand the propagators in supersoft, IR, scalar masses before integration.

Hence, scalar propagators are treated as massless and their mass effect
is included as a perturbative two-point interaction,
presented as a blob at N$^4$LO in eq.~\eqref{eq:ss:tree}.
These two-point interaction vertices appear at two-loop level, and are effectively
counted as three-loop diagrams.
The field normalization factor $Z_s$,
resulting from
the soft gauge field modes, is computed at one-loop order;
see e.g.~\cite{Hirvonen:2021zej}.
Pure scalar-field diagrams vanish in dimensional regularization
given that no scales appear in the propagators.
They do not contribute to the matching.
However,
scalar-field loops are computed within the supersoft EFT in the IR,
and at one-loop order are
\begin{align}
  \label{eq:ss:1loop}
  S_{\rmi{supersoft}}^{\rmi{1-loop}} &=
  \underbrace{
  	\TopoVR(\Asr1)
  }_{\text{N$^2$LO}}
  + \underbrace{
  	\TopoVRo(\Asr1)
  	\TopoVRb(\Asr1)
  }_{\text{N$^4$LO}}
  \,,
\end{align}
where
the ``$x$-expansion'' is enforced by treating higher-order contributions
(in powers of $x$) to the scalar mass as perturbative corrections.
These corrections are denoted by a blob  insertion of $\Pi_\phi$;
see eq.~\eqref{eq:mass-insertion}.
A box denotes the insertion of the two-point vertex $\bar{\Pi}_\phi$
arising from field renormalization;
see eq.~\eqref{eq:kinetic-insertion}.
Two-loop diagrams within the supersoft EFT are of higher order than N$^4$LO and
can thus be omitted in our computation.
Combining the results for
the soft expansion of eq.~\eqref{eq:ss:tree} and
the supersoft corrections in eq.~\eqref{eq:ss:1loop}, yields
the perturbative expansion of
the broken-phase effective potential
in fig.~\ref{fig:diags:Veff}.

\section{\\
  Thermodynamics of U(1) + Higgs
  theory
  }
\label{sec:u1:results}

The three-dimensional theory action
of the Abelian Higgs model is given by
eq.~\eqref{eq:S3d} and its
tree-level
potential by eq.~\eqref{eq:V:3d}
where the scalar $\Phi$ transforms under ${\rm U}(1)$.
This model has previously been studied
in~\cite{Karjalainen:1996wx,Karjalainen:1996rk,Kajantie:1997vc,Kajantie:1997hn,Kajantie:2004vy,Kajantie:2001ts}.
Contrary to the ${\rm SU}(2)$ model,
there is no magnetic mass for the photon.
Thus making four-loop computations tractable;
at least in principle.
However, a non-perturbative mass exists for
finite lattice spacings~\cite{Banks:1977cc,Ambjorn:1982ts}, and
topological effects such as vortices
are relevant~\cite{Kajantie:1998bg,Kajantie:1998zn,Kajantie:1999ih}.
In the context of ${\rm SU}(2)$ + adjoint Higgs theory,
similar effects were recently discussed in~\cite{Niemi:2022bjg}.

For the symmetric phase,
the first two orders
$
\Veff\bigr|_{\rmii{LO}}^{\rmi{sym}}=
\Veff\bigr|_{\rmii{NLO}}^{\rmi{sym}} \equiv 0$.
The first non-vanishing contributions arise at
N$^3$LO and
N$^4$LO,
\begin{align}
  \Veff\bigr|_{\rmii{N$^2$LO}}^{\rmi{sym}} &=
  -\frac{1}{6\pi}
    \bigl(m_3^2\bigr)^{\frac{3}{2}}
  \,, \\
  \Veff\bigr|_{\rmii{N$^3$LO}}^{\rmi{sym}} &=
  -\frac{g_3^{2} m_3^2}{(4\pi)^2} \frac{1}{2}\Bigl[
  - 3
  + 4\ln2
  + 2\ln\frac{m_3^2}{\Lamd^2}
  \Bigr]
  \,,\\
  \Veff\bigr|_{\rmii{N$^4$LO}}^{\rmi{sym}} &=
  \frac{g_3^{4}}{(4\pi)^3}
  \frac{\bigl(m_3^2\bigr)^{\frac{1}{2}}}{12}
  \Bigl[
  97
  + 4\pi^2
  \nn &
  \hphantom{\frac{g_3^{4}}{(4\pi)^3}
  \frac{\bigl(m_3^2\bigr)^{\frac{1}{2}}}{12}\Bigl[}
  - 136\ln2
  - 24\ln\frac{m_3^2}{\Lamd^2}
  \Bigr]
  \,.
\end{align}

The broken-phase effective potential amounts to
\begin{align}
  \Veff\bigr|_{\rmii{LO}}^{\rmi{bro}} &=
  \frac{1}{2} m_{3}^{2} \phi^2
  + \frac{1}{4} \lambda_{3}^{ } \phi^4
  - \frac{1}{6\pi} g_{3}^{3} \phi^3
  \,, \\
  \Veff\bigr|_{\rmii{NLO}}^{\rmi{bro}} &=
  \frac{g_3^{4}\phi^2}{(4\pi)^2}\Bigl[
  - 1
  - 2\ln\frac{\Lamd}{2g_{3}\phi} \Bigl]
  \,, \\
  \Veff\bigr|_{\rmii{N$^2$LO}}^{\rmi{bro}} &=
  -\frac{1}{12\pi}\Bigl[
  \widetilde{m}_{h}^3
  + \widetilde{m}_{\rmii{$G$}}^{3}
  \Bigl]
  \,,\\
  \Veff\bigr|_{\rmii{N$^3$LO}}^{\rmi{bro}} &=
  \frac{4 g_{3}^{2}}{(16\pi)^2}\Bigl[
  m_{3}^{2}\Bigl(
  3
  + 4\ln2
  + 8 \ln\frac{\Lamd}{2g_{3}\phi}
  \Bigr)
  \nn &
  \hphantom{\frac{4 g_{3}^{2}}{(16\pi)^2}}
  + \lambda_{3}^{ }\phi^2\Bigl(
  7
  + 4\ln2
  + 16\ln\frac{\Lamd}{2g_{3}\phi}
  \Bigr)
  \Bigr]
  \nn &
  + \frac{g_{3}^{5}\phi}{(16\pi)^3}\frac{16}{3}\Bigl[
  77
  + 4\pi^2
  - 104\ln2
  \Bigr]
  \,,\\[2mm]
  \Veff\bigr|_{\rmii{N$^4$LO}}^{\rmi{bro}} &=
  - \frac{g_{3}^{ }}{(16\pi)^2}\frac{1}{\phi}\frac{4}{3}\Bigl[
  11 \widetilde{m}_{h}^3
  + 16 \widetilde{m}_{\rmii{$G$}}^3
  \Bigr]
  \nn &
  + \frac{2g_{3}^{4}}{(4\pi)^3}\Bigl[
  -\widetilde{m}_{h}^{ }
  + (
      \widetilde{m}_{h}^{ }
    + \widetilde{m}_{\rmii{$G$}}^{ }
    )
  \ln\frac{\Lamd}{2g_{3}\phi}
  \Bigr]
  \,.
\end{align}

The perturbative results
of the critical mass and condensates
for this model up to N$^2$LO, are
\begin{align}
\label{eq:mcRes:AH}
  \yc &=
  \frac{1}{2^{ }(3\pi)^{2} x^{ }}\Bigl[
  1
  + \frac{9}{2} x \ln\tilde{\mu}_{3}
  - \Bigl(\frac{x}{2}\Bigr)^{\frac{3}{2}}
  \Bigr]
  \,,\\[1mm]
  \label{eq:phisqRes:AH}
  \Delta\bigl\langle \hsq \bigr\rangle_\rmi{c} &=
  \frac{1}{2^{ }(3\pi)^{2} x^{2}}\Bigl[
  1
  - \frac{9}{2} x
  + \frac{25}{2} \Bigl(\frac{x}{2}\Bigr)^{\frac{3}{2}}
  \Bigr]
  \,,\\[2mm]
  \label{eq:phisqsqRes:AH}
  \Delta\bigl\langle \left(\hsq\right)^2 \bigr\rangle_\rmi{c} &=
  \frac{1}{2^{2}(3\pi)^{4} x^{4}}\Bigl[
  1
  - 9 x
  + 13\Bigl(\frac{x}{2}\Bigr)^{\frac{3}{2}}
  \Bigr]
  \,,
\end{align}
where
$\tilde{\mu}_3\equiv
e^{\tfrac{1}{2}+\ln\tfrac{3}{2}} (\pi x\Lamd)$.
After including three-loop corrections,
we find the critical mass and condensates at
N$^3$LO and
N$^4$LO to be
\begin{widetext}
\begin{align}
  \yc\bigr|_{\rmii{N$^3$LO}}
  &= - \frac{x^{2}}{{2^{3}(4\pi)^2 x^{ }}}
  \biggl[
  64\ln\bigl(x^\frac{9}{8}\Lamd\pi\bigr)
  + 63
  + 4\pi^{2}
  - 152\ln2
  + 64\ln3
  \biggr]
  \,,\\[2mm]
  \Delta\bigl\langle\hsq\bigr\rangle_\rmi{c}\bigr\vert_{\rmii{N$^3$LO}}
  &=\frac{x^{2}}{{2^{3}(4\pi)^{2} x^{2}}}
  \biggl[
  24\ln x
  + 111
  + 4\pi^{2}
  - 104\ln2
  \biggr]
  \,,\\[2mm]
  \Delta\bigl\langle\bigl(\hsq\bigr)^2\bigr\rangle_\rmi{c}\bigr\vert_{\rmii{N$^3$LO}} &=
  \frac{x^{2}}{{2^{6}(3\pi)^4 x^{4}}}
  \biggl[
  216\ln x
  + \frac{1755}{2}
  + 18\pi^{2}
  - 576\ln2
  + 144\ln(3\pi)
  \biggr]
  \,,\\[2mm]
  \yc\bigr|_{\rmii{N$^4$LO}} &=
  - \frac{x^{\frac{5}{2}}}{{2^{\frac{5}{2}}(4\pi)^2 x^{ }}}\Bigl[
    12\ln x
    - \frac{85}{6}
    - 2\pi^{2}
    + 32\ln2
  \Bigr]
  \,,\\[2mm]
  \Delta\bigl\langle\hsq\bigr\rangle_\rmi{c}\bigr\vert_{\rmii{N$^4$LO}} &=
  \frac{x^{\frac{5}{2}}}{{2^{\frac{7}{2}} (4\pi)^2 x^{2}}}
  \Bigl[
    60\ln x
    - \frac{143}{2}
    - 10\pi^{2}
    + 160\ln2
  \Bigr]
  \,,\\[2mm]
  \Delta\bigl\langle\bigl(\hsq\bigr)^2\bigr\rangle_\rmi{c}\bigr\vert_{\rmii{N$^4$LO}}
  &= \frac{x^{\frac{5}{2}}}{{2^{\frac{9}{2}} (3\pi)^4 x^{4}}}
  \Bigl[
  54\ln x
  - \frac{1341}{8}
  - 9\pi^{2}
  + 144\ln2
  \Bigr]
  \,.
\end{align}
\end{widetext}

\section{%
  Thermal EFT for SU(2) with a doublet and singlet dark sector}
\label{sec:EFT:su2}

This appendix details the model employed in sec.~\ref{sec:GW}.
While most of the results can be extracted from
the literature~\cite{Kajantie:1995dw,Brauner:2016fla,Schicho:2021gca,Niemi:2021qvp,Tenkanen:2022tly},
we include and display novel thermal corrections.

\subsection{Model at zero temperature}
\label{sec:4D:su2+doublet+singlet}

We define our setup through the Lagrangian
\begin{align}
\label{eq:lag-4d}
  \mathcal{L} =
  \frac{1}{4} F_{\mu\nu}^a F_{\mu\nu}^a
  &+ (D_\mu \Phi)^\dagger (D_\mu \Phi)
  \nn &
  + \frac12 (\partial_{\mu} S)^2
  + V(\Phi,S)
  \,,
\end{align}
where the covariant derivative
$D_\mu \Phi = (\partial_\mu - i g A_\mu)\Phi$,
where $A_\mu = A_\mu^a T^a$.
The scalar potential is
\begin{align}
  V(\Phi,S) &=
  m^2_\phi \hsq
  + \lambda_\phi^{ } (\hsq)^2
  + \frac12 m_{s}^2 S^2
  + \frac14 \lambda_{s}^{ } S^4
  \nn &
  + \frac12 \mu_{m}^{ } (\hsq)^2 S
  + \frac12 \lambda_{m}^{ } (\hsq)^2 S^2
  \nn &
  + \mu_1 S
  + \frac13 \mu_3 S^3
  \,.
\end{align}
The cubic portal
allows for doublet-singlet mixing.
We work in a parametrization where
the mixing angle $\sin \theta$,
the scalar mass squared eigenvalues
$M^2_\phi$,
$M^2_s$,%
\footnote{
  For a small mixing angle
  $\sin \theta \ll 1$,
  the mass
  $M_\phi$ describes the {\em mostly} doublet and
  $M_s$ the {\em mostly} singlet state.
}
the doublet vacuum expectation value (VEV) at zero temperature $v_0$,
the gauge coupling $g^2$,
the singlet self interactions
$\lambda_s$,
$\mu_3$, and
the portal coupling $\lambda_m$
are treated as input parameters.
We fix the singlet VEV to vanish which
fixes singlet tadpole coupling
$\mu_1 = -\mu_m v^2_0/4$.
Concretely, we use
the tree-level relations
\begin{align}
m^2_\phi &= -\frac14 \Big( M^2_\phi + M^2_s + (M^2_\phi - M^2_s) \cos(2\theta) \Big)
  \,,\nn
  m^2_s &= \frac12 \Big( -\lambda_m^{ } v^2_0 +  M^2_\phi + M^2_s - (M^2_\phi - M^2_s) \cos(2\theta) \Big)
  \,,\nn
\lambda_\phi &=
    \frac{1}{4v^2_0} \Big( M^2_\phi + M^2_s + (M^2_\phi - M^2_s) \cos(2\theta) \Big)
  \,, \nn
\mu_m &= -\frac{2}{v_0} (M^2_\phi - M^2_s) \sin \theta \cos\theta
  \,.
\end{align}
For a realistic theory,
instead of a dark-sector toy setup,
it would be warranted to improve these relations by
the corresponding zero-temperature one-loop
corrections~\cite{Kajantie:1995dw}, cf.~\cite{Niemi:2021qvp}.

\subsection{Effective theory}

At high temperatures $m^2_\phi/T, m^2_s/T \sim g \ll 1$
and
the 3D zero Matsubara modes live
in the IR at the soft scale $|p|\sim\mathcal{O}(g T)$.
In turn,
the non-zero Matsubara modes in the UV at scale $|p| \sim \mathcal{O}(\pi T)$
can be integrated out~\cite{Laine:2016hma}.
The resulting 3D~EFT
is of the same functional form as eq.~\eqref{eq:lag-4d} with
the addition of
the Lagrangian for
the temporal component of
the gauge field
\begin{align}
\label{eq:lag-temporal}
\mathcal{L}^{\rmii{3d}}_\rmi{temporal}&=
    \frac{1}{2}(D_{r}^{ }A^a_0)^2
  + \frac{1}{2} \mD^{2}\,A^a_0A^a_0
  \nn &
  + \frac{1}{4}\kappa_{3}^{ }\,(A^a_0A^a_0)^2
  + \frac{1}{2} h_3^{ }\,\he\phi\phi A^a_0A^a_0
  \nn &
  + \frac{1}{4} y_{3}^{ }\,S^2 A^a_0A^a_0
  + \frac{1}{2} x_{3}^{ }\,S^2 A^a_0A^a_0
  \;,
\end{align}
where the covariant derivative
for triplet ${\rm SU}(2)$ temporal scalars, reads
$D_{r}^{ }A_{0}^{a} = \partial_{r}^{ } A^a_0+ g_{3}^{ } \epsilon^{abc}A^b_rA^c_0$.
Here,
we chose different normalizations for
$h_3$~\cite{Kajantie:1995dw,Croon:2020cgk},
$y_3$, and
$x_3$ compared to~\cite{Schicho:2021gca}.
In addition,
$\kappa_3 \equiv 2\text{\tt $\lambda$VLL[1]}$ compared to
the convention used by {\tt DRalgo};
see the
{\tt dark-su2-higgs-singlet.m}
model file~\cite{SU2HiggsSinglet}.

In analogy to eq.~\eqref{eq:ss:tree},
the construction of the EFT can be
schematically illustrated in terms of
the effective action as
\begin{align}
\label{eq:DR:action}
  &S_{\rmii{EFT}}[\varphi] =
  \int_{\vec{x}} \frac{1}{2} (\partial_i \varphi)^2 Z_\varphi
+ \Vtxz
+ \TopoVR(\Asa1)
+ \ToptVS(\Asa1,\Asa1,\Lsa1)
  \nn[2mm]
&=\!
\int_{\vec{x}} \bigg\{
p_0^{\varphi=0}
+ \bigg[
    \Vtxt(\Lsrxi,\Lsr1)
  + \TopSii(\Lsrxi,\Lsri,1)
  + \TopSii(\Lsri,\Lsri,2)
  \bigg]
  \nn[1mm] &
  \hphantom{{}=\int_{\vec{x}} \bigg\{p_0^{\varphi=0}}
  \!+ \bigg[
    \Vtxv(\Lsrxi,\Lsri,\Lsri,\Lsri)
  + \TopVi(\Lsr1,\Lsr1,\Lsr1,\Lsr1,1)
  \bigg]
  + \bigg[
    \TopVIi(\Lsr1,\Lsr1,\Lsr1,\Lsr1,\Lsr1,\Lsr1,1)
  \bigg] \bigg\}
  \nn[2mm]
  &
  =\!
  \int_{\vec{x}} \bigg\{
  \underbrace{\!
  p_0^{\varphi=0}
  + \frac{1}{2} \varphi^2 \big( m^2_3 \big)
  + \frac{1}{4} \varphi^4 \big( \lambda_3 \big)}_{\mathcal{O}(g^4)}
  + \mathcal{O}(g^6 \varphi^6) \bigg\}
  \,,
\end{align}
where
encircled numbers indicate the loop order and
boxes indicate kinetic insertions on external lines
in analogy with eq.~\eqref{eq:kinetic:insert}.
Here, the effective action is computed using
the background field method, in terms of
a formal scalar background field $\varphi$.
For simplicity,
we present here only one such background field, keeping in mind that
for multiple scalar fields within the EFT, each field has its own background.
Furthermore, in this simplified illustration,
we do not detail the background field method for the gauge sector.

In the first line of eq.~\eqref{eq:DR:action},
loop diagrams involving Matsubara sum-integrals are integrated over
the hard momenta, i.e.\
over non-zero Matsubara modes in the UV.
Masses $M^2(\varphi)$ in propagators depend on the background field, and
sum-integrals are computed in
the high-temperature expansion of $M^2/T^2 \sim g^2$.
Matsubara zero modes in the IR are treated as massless.
The field normalization factor
\begin{align}
  Z_\varphi = 1 + \frac{{\rm d}}{{\rm d}k^2}
  \biggl[\TopoSB(\Lsri,\Asai,\Asai)\biggr]
  \,,
\end{align}
is likewise obtained by integrating over
non-zero Matsubara modes, and
computing
the relevant scalar two-point functions in
an expansion of soft external momenta $k$.

In the second line of eq.~\eqref{eq:DR:action},
we have expanded the action in terms of background field,
$\varphi$.
We have highlighted that the unit operator $p_0$ is field-independent, and
return to its computation below.
Terms with non-vanishing background field correspond to
Green's functions generated by the effective action, and
the ellipsis denotes
eight- and higher-point correlators.

In the third line of eq.~\eqref{eq:DR:action},
we identify these Green's functions
with EFT parameters of the scalar potential,
that we denote by
a schematic
$m^2_3$ and
$\lambda_3^{ }$ for this simplified illustration.
We remark,
that in the second line,
the effect of
the $Z$-factor is captured
by the Green's functions%
\footnote{
  $Z$-factor contributions are crucial to obtain
  a renormalization-scale invariant, gauge-independent result~\cite{Croon:2020cgk,Hirvonen:2021zej}.
}
which is illustrated by a box on the external lines.
In practice,
the 3D parameter matching relations can be determined by
directly computing Green's functions without the presence of the background field,
instead of computing the action.
This is practical,
especially for
the gauge, and gauge-scalar mixed sectors~\cite{Kajantie:1995dw}.

Assuming a power counting in which
the parent theory
squared mass parameters are soft $\sim(gT)^2$, in
the high-$T$ expansion and
the quartic couplings scale as $\mathcal{O}(g^2)$,
allows for determining all 3D~EFT parameters at
$\mathcal{O}(g^4)$ provided that
the masses are determined at
two-loop and couplings at one-loop level.
Sextic and higher operators can be neglected
as they contribute at $\mathcal{O}(g^6)$.

Within the so-far constructed EFT,
the Debye mass $\mD^2$ is soft, and parametrically larger than
the mass of the scalar doublet $\Phi$ that undergoes the transition deeper
in the IR.
This allows to integrate out
the temporal scalar $A^a_0$, to build an EFT at
a {\em softer} scale.
In addition, we assume that
the singlet mass
$m_s \sim \mD$, and hence
the singlet is integrated out as well.
The resulting EFT in the IR is given by eq.~\eqref{eq:S3d}.

Constructing this softer EFT aligns with eq.~\eqref{eq:DR:action} with
a few modifications.
In this case, only
the soft loops of the temporal scalar and
the singlet are integrated over, while
the doublet (as well as spatial gauge field) are treated as massless.
The convergence of the soft theory is slower as
each new loop order is suppressed by $g$,
whereas in
the hard-to-soft matching, the loop suppression is $g^2$.
Typically,
in softer theory only one-loop effects from
the soft scale are included for
the scalar quartic coupling~\cite{Kajantie:1995dw}.
While this is numerically an excellent approximation for
contributions of the temporal scalar,
for the singlet, also two-loop effects can be sizable.
Similarly, effects from
the soft singlet can induce sizable corrections to
the sextic operator,
to which soft contributions in general are parametrically
$\mathcal{O}(g^3)$~\cite{Kajantie:1995dw}.
While at first glance this might seem alarming,
we comment on these issues in sec.~\ref{sec:higher:dim}
when we present concrete expressions and further discuss
our numerical analysis.

The unit operator in the final EFT composes of
$p_0 = p_{\rmii{E}} + p_\rmii{M}$~\cite{Gynther:2005av},
where hard and soft contributions,
typically labelled as
electrostatic ($p_{\rmii{E}}$) and
magnetic ($p_{\rmii{M}}$),
follow from vacuum diagrams of the form (here in four-dimensional units)
\begin{align}
\label{eq:pE:schematic}
p_{\rmii{E}} =
\TopoVR(\Asai)
  &
  + \ToptVS(\Asai,\Asai,\Lsai)
  + \ToprVW(\Asai,\Asai,\Asai,\Lsai,\Lsai,\Lsai)
  \nn[1mm] &
\sim T^4 \Bigl( g^0 + g^2 + g^4 + \mathcal{O}(g^6) \Bigr)
  \,, \\[2mm]
\label{eq:pM:schematic}
p_{\rmii{M}} =
T \Bigl(
\TopoVR(\Asr1)
  &
  + \ToptVS(\Asr1,\Asr1,\Lsr1)
  + \ToprVW(\Asr1,\Asr1,\Asr1,\Lsr1,\Lsr1,\Lsr1) \Bigl)
  \nn[1mm] &
  \sim T^4 \Bigl( g^3 + g^4 + g^5 + \mathcal{O}(g^6) \Bigr)
\,,
\end{align}
where
each loop order is represented by its most complicated topology.

For
$p_\rmii{E}$,
the loop integration is over non-zero Matsubara modes and
each order is suppressed by $g^2$.
For
$p_\rmii{M}$,
loops involve soft fields and each order is suppressed by $g$.
In both cases, all contributions are purely perturbative and
for hot QCD,
the computation of
the four-loop $\mathcal{O}(g^6)$
has recently been organized at the level of master integrals~\cite{Navarrete:2022adz}.

\subsection{Collection of expressions}

This section collects all relevant thermal corrections for the softer EFT.
Installing the power counting
\begin{align}
  m^2_\phi, m^2_s &\sim (gT)^2
  \,,&
  \lambda_\phi, \lambda_s, \lambda_m &\sim g^2
  \,,
  \nn &&
  \mu_3, \mu_m &\sim g^2 T
  \,,
\end{align}
leads to a similar EFT construction as in~\cite{Niemi:2021qvp},
and the hard-to-soft matching relations can be read from therein
(by omitting
${\rm U}(1)$,
${\rm SU}(3)$, and
fermionic sectors contributions).
Alternatively, one can compute the parameters using
{\tt DRalgo}~\cite{Ekstedt:2022bff}
via the model file~\cite{SU2HiggsSinglet}.
Hence, we do not list these relations here.

The matching
results in the
two-loop mass parameters (and singlet tadpole) and
one-loop couplings.
Such relations are renormalization-scale invariant at
$\mathcal{O}(g^4)$,
and in our numerical analysis,
we apply one-loop $\beta$-functions to run parameters to
the optimal thermal scale
$\LamD_\rmi{opt} = 4\pi e^{-\gammaE} T$~\cite{Farakos:1994kx},
where $\gammaE$ is the Euler-Mascheroni constant.
Since the required $\beta$-functions can be obtained with
{\tt DRalgo}, we do not list them explicitly.

If the singlet decouples,
we need smaller doublet self-coupling which satisfies
the parametric scaling
$\lambda_{\phi} \sim g^3/\pi$;
cf.~\eqref{eq:BMA}.
Also in this case,
we include $\mathcal{O}(\lambda^2_\phi)$ terms in
the hard-to-soft matching,
yet remark that these contributions,
as well as $\mathcal{O}(g^2 \lambda_\phi)$, are
miniscule compared
to pure gauge contributions of $\mathcal{O}(g^4)$ that dominate.

In~\eqref{eq:BMB},
and by further fixing $\lambda_m = 2.5$ in this section,
we find that
$\lambda_\phi$,
$\lambda_m$,
$\lambda_s$, and
the gauge coupling contribute with approximately equal importance
to the two-loop thermal mass parameters,
while the effect from cubic couplings is less important.%
\footnote{
  Contributions of
  $\mu_3$,
  $\mu_m$
  to one-loop thermal masses
  are $\propto L_b$ and
  vanish at the optimized $\LamD_\rmii{opt} = 4\pi e^{-\gammaE} T$.
  Further corrections to the matching relations from these couplings are
  $\propto \zeta_3/(4\pi)^4$ or
  $\zeta_5/(4\pi)^6$
  and we find them to be negligible;
  cf.\ sec.~4.2 of~\cite{Schicho:2021gca}.
}
Cubic couplings, however, yield the largest (one-loop) correction for
the singlet tadpole which is dominated by
its tree-level value while two-loop corrections are further suppressed.
For both mass parameters, tree-level and one-loop contributions are dominant while
the two-loop correction is subdominant.
This indicates a controlled high-$T$ expansion, and
concretely
$m^2_\phi/T \sim 1/4$ and
$m^2_s/T \sim 4/5$ in vicinity of critical temperature $\Tc \sim 188$~GeV.

The unit operator for
the hard modes from eq.~\eqref{eq:pE:schematic} and
the soft modes from eq.~\eqref{eq:pM:schematic},
can be obtained with {\tt DRalgo} and reads
\begin{widetext}
\begin{align}
\label{eq:pE}
p_{\rmii{E}} &=
    11 \frac{\pi^2}{90} T^4 
  - \frac{T^2}{576} \Big(
      T^2 (39 g^2 + 24 \lambda_\phi + 4 \lambda_m + 3 \lambda_s)
    + 24 (4 m^2_\phi + m^2_s )
    \Big)
    \nn &
    + \frac{T^2}{(4\pi)^2} \Big[
      \frac{m^2_s}{24} L_b \Bigl(2 \lambda_m + 3 \lambda_s\Bigr)
      + m^2_\phi \Big(
          L_b \Bigl(\lambda_\phi + \frac{1}{12} \lambda_m\Bigr)
        + \frac{g^2}{8} \Bigl(-2 + 12\ln(2\pi) - 12(\ln\zeta_2)' + 9 L_b\Bigr)
        \Big)
    \nn &
    \hphantom{{}\frac{T^2}{(4\pi)^2} \bigg[}
    - \frac{1}{48}\bigl(4 \mu^2_3 + 3 \mu^2_m\bigr)
      \bigl(2\ln(2\pi) - 2(\ln\zeta_2)' + L_b\bigr)
    \Big] 
    -  \frac{L_b}{(4\pi)^2}
        \bigl(\mu^4_\phi + \frac{1}{4} \mu^4_s\bigr) 
    \nn &
    +  \frac{T^4}{(4\pi)^2} \Big[
      \alpha_{\rmii{E$g^4$}} g^4
    \!+\! \alpha_{\rmii{E$\lambda_\phi g^2$}} \lambda_{\phi}^{ } g^2
    \!+ \alpha_{\rmii{E$\lambda^2_\phi$}} \lambda_{\phi}^{2}
    \!+ \alpha_{\rmii{E$\lambda_m g^2$}} \lambda_{m}^{ } g^2
    \!+ \alpha_{\rmii{E$\lambda_\phi \lambda_m$}} \lambda_{\phi}^{ }\lambda_{m}^{ }
    \!+ \alpha_{\rmii{E$\lambda^2_m$}} \lambda_m^{2}
    \!+ \alpha_{\rmii{E$\lambda_m \lambda_s$}} \lambda_{m}^{ } \lambda_{s}^{ }
    \!+ \alpha_{\rmii{E$\lambda^2_s$}} \lambda_{s}^{2}
  \Big] 
  \,,\\[2mm]
\label{eq:pM}
\frac{p_{\rmii{M}}}{T} &=
    -\frac{1}{12\pi} \Big( 3\mD^3 + m^3_{s,3} \Big)
    - \frac{1}{(4\pi)^2} \Big[
          g^2_3 \mD^2 \Big(\frac{9}{2} + 6 \ln \frac{\Lamd}{2\mD} \Big)
      + \frac{3}{4} \Big(5 \kappa_3^{ } \mD^2 + \lambda_{s,3}^{ } m^2_{s,3}\Big)
    \Big]
    \nn &
    - \frac{2\mD}{(4\pi)^3} \Big[
        g^4_3 \bigl(23 + \pi^2 - 11 \ln 2\bigr) + 12 h^2_3
      + \frac{9}{8}\bigl(g^4_3 + 8 h^2_3\bigr) \ln\frac{\Lamd}{2\mD}
  \Big]
  \,, 
\end{align}
\end{widetext}
where
$L_b \equiv 2\ln\frac{\LamD e^\gammaE}{4\pi T}$,
$\zeta_n = \zeta(n)$ is the Riemann zeta function, and
$(\ln\zeta_n)' \equiv \zeta'(n)/\zeta(n)$.
For the $T^4$-coefficients, we adopted the (slightly modified) notation
of~\cite{Gynther:2005dj,Tenkanen:2022tly},
and results for these coefficients can also be read from therein.
In fact,
the term proportional to $\kappa_3$ in eq.~\eqref{eq:pM} is
$\mathcal{O}(g^6 T^4)$ and could hence be dropped.
In
the three-loop contribution, however, we have included
only the temporal scalar contribution,
as these can be conveniently read
from~\cite{Gynther:2005av},
but not
the singlet contributions which are currently
beyond reach for the fangs of {\tt DRalgo}.

\begin{table}[t]
\begin{tabular}{|l|c|c|}
  \hline
  \ref{eq:BMA} &
  hard &
  soft
  \\ \hline\hline
  1-loop &
  $-$
  &
  1.60\%
  \\
  2-loop &
  1.7\%
  &
  0.32\%
  \\
  3-loop &
  0.9\%
  &
  0.30\%
  \\ \hline
\end{tabular}
\hspace{1mm}
\begin{tabular}{|l|c|c|}
  \hline
  \ref{eq:BMB} &
  hard &
  soft
  \\ \hline\hline
  1-loop &
  $-$
  &
  \,8.7\%\hphantom{$^{a}$}
  \\
  2-loop &
  6.9\%
  &
  \,1.7\%\hphantom{$^{a}$}
  \\
  3-loop &
  3.3\%
  &
  \,0.1\%%
  \footnote{
    This number is only for temporal mode contributions
    as singlet terms were omitted as described earlier.
  }
  \\ \hline
\end{tabular}
  \caption{%
    Corrections to the unit operator
    $p_0 = p_{\rmii{E}} + p_\rmii{M}$ with respect to
    its LO result
    for benchmark points~\eqref{eq:BMA} and \eqref{eq:BMB}
    originating from soft and hard modes.
    }
  \label{tab:p0:corrections}
\end{table}

Corrections from different loop orders to
$p_0 = p_{\rmii{E}} + p_\rmii{M}$ with respect to its LO result are
listed in tab.~\ref{tab:p0:corrections}.
For~\eqref{eq:BMA}, and as already highlighted above,
these corrections are dominated by pure gauge contributions,
and corrections involving
$\lambda_\phi^{ }$ and
$m^2_\phi/T^2$ are minor.
Also for~\eqref{eq:BMB}, these relative corrections
display great convergence, albeit not as good as in~\eqref{eq:BMA}.

The explicit ``soft-to-softer'' EFT matching relations,
where the temporal scalar and the singlet are integrated out,
read
\begin{align}
\label{eq:soft-DR}
\bar{g}^2_3 &= g^2_3
    - \bigg[\frac{1}{24\pi}\frac{g^4_3}{\mD} \bigg]
    \,, \\
\label{eq:lam-softer}
\bar{\lambda}_{\phi,3} &=
    \lambda_{\phi,3}
    - \bigg[
    \frac{1}{32\pi}\Big( \frac{3 h^2_3}{\mD}
    + \frac{\lambda^2_{m,3}}{m_{s,3}} \Big)
    \bigg]
  \nn &
    +
    \frac{1}{(4\pi)^2} \bigg[
        \frac{1}{\mD^2} \Big(
          \frac34 h^3_3
        - \frac32 h^2_3 g^2_3
        + \frac38 h_3^{ } g^4_3
        - \frac{3}{128} g^6_3
        \Big)
  \nn &
    \hphantom{{}+\frac{1}{(4\pi)^2} \bigg[}
      + \frac{1}{4} \frac{\lambda^3_{s,3}}{m^2_{s,3}}
    \bigg]
  + \Gamma_{3}
  \,,\\
\label{eq:mass-softer}
\bar{m}^2_{\phi,3} &=
    m^2_{\phi,3}
    - \bigg[\frac{1}{8\pi}\Big( 3 h_3 \mD + \lambda_{m,3} m_{s,3} \Big) \bigg]
    \nn &
    - \frac{1}{(4\pi)^2} \bigg[
        \frac34 h^2_3
      - \frac{15}{4} h_3^{ } \kappa_3^{ }
      - \frac32 h_3^{ } g^2_3
      - \frac34 \lambda_{m,3}^{ } \lambda_{s,3}^{ }
    \nn &
    \hphantom{{}\frac{1}{(4\pi)^2} \bigg[}
      + \Big(
          \frac32 h^2_3
        - 6 h_3^{ } g^2_3
        + \frac34 g^4_3
      \Big) \ln\frac{\Lamd}{2\mD}
    \nn &
    \hphantom{{}\frac{1}{(4\pi)^2} \bigg[}
    + \frac14 \lambda^2_{m,3}\Bigl(
        1
      + 2\ln\frac{\Lamd}{2 m_{s,3}}
      \Bigr)
    \bigg]
    + \Pi_{3}
    \,.
\end{align}
Here,
we have included two-loop contributions to
the self-coupling.
While these corrections are often omitted,
they are formally of the same order as
the two-loop thermal mass.
They can readily be found using
the two-loop effective potentials computed in~\cite{Niemi:2021qvp,Niemi:2020hto}.

The contributions of
the singlet cubic couplings are
\begin{align}
\label{eq:Gamma3-cubic}
\Gamma_{3} &=
      \frac{1}{4} \mu_{1,3}^{ } \Big(
        2\frac{\lambda_{m,3}^{ } \mu_{m,3}^{ }}{m^4_{s,3}}
      - \frac{\mu_{3,3}^{ } \mu^2_{m,3}}{m^6_{s,3}} \Big)
    - \frac{\mu^2_{m,3}}{8 m^2_{s,3}}
    \nn &
    + \frac{\mu_{m,3}^{2}}{32 \pi m^5_{s,3}} \Big(
        \frac{5}{4} \mu_{m,3}^{ }
      - \mu_{m,3}^{ } \mu_{3,3}^{ }
      - \mu_{3,3}^{2}
      \Big)
    \nn &
    + \frac{\mu_{m,3}^{2}}{32 \pi m^3_{s,3}} \Big(
      5 \lambda_{m,3} - 12 \lambda_{\phi,3} -3 \lambda_{s,3}
    -2 \frac{\lambda_{m,3} \mu_{3,3}}{\mu_{m,3}} \Big)
    \nn &
    -\frac{1}{48 (4\pi)^2 m^4_{s,3}} \Big( 4 \mu^2_{3,3} + 3 \mu^2_{m,3} \Big) 
- \frac{\lambda_{\phi,3}}{24\pi} \frac{\mu^2_{m,3}}{m^3_{s,3}}  
  \,, \\[2mm]
\label{eq:Pi3-cubic}
\Pi_{3} &=
    -\frac{\mu_{1,3} \mu_{m,3}}{2 m^2_{s,3}}
+ \frac{\mu_{m,3}}{16\pi m_{s,3}} \big( 2 \mu_{3,3} - \mu_{m,3} \big)
    \nn &
    + \frac{\mu_{1,3}}{32\pi} \bigg[
        \frac{\mu^2_{m,3}}{m^5_{s,3}} \big( \mu_{m,3}\!- 2 \mu_{3,3} \big)
      + \frac{4\lambda_{m,3}^{ }}{m^3_{s,3}} \big(2\mu_{m,3}\!+ \mu_{3,3} \big) \bigg]
    \nn &
+ \frac{\lambda_{m,3}}{24(4\pi)^2 m^2_{s,3}}
      \big( 4 \mu^2_{3,3} + 3 \mu^2_{m,3} \big)
- \frac{m^2_{\phi,3}}{48\pi} \frac{\mu^2_{m,3}}{m^3_{s,3}}  
    \,.
\end{align}
Our computation of these contributions
is diagrammatic~\cite{Brauner:2016fla}
and involves
a class of one-particle-reducible (1PR) diagrams as detailed therein.
We, however, integrate out the singlet at the soft scale,
whereas in~\cite{Brauner:2016fla} the singlet is
integrated out along with non-zero Matsubara modes.
Consequently the integrals associated with
the required diagrams are different.
We have included the 1PR contributions at
tree- and one-loop level but omitted them at two-loop level.
There, we only include
one-particle-irreducible (1PI) contributions.
These 1PI contributions
can be found using
the two-loop effective potential~\cite{Niemi:2021qvp,Niemi:2020hto}.
In addition, we have included the leading contribution from
the one-loop $Z$-factor induced by the singlet,
described by the very last terms in
eqs.~\eqref{eq:Gamma3-cubic} and \eqref{eq:Pi3-cubic}.

In~\eqref{eq:BMB}, in which
$\mD/T \sim 4/5$ and
$m_{s,3}/T \sim 4/3$ in vicinity of $\Tc$ and
where integrating out the singlet is justified,
we find that
the dominant effect in
$\Gamma_3$ and
$\Pi_3$ comes from the tree-level terms.
The one-loop contributions $\propto 1/\pi$ are subdominant while
two-loop as well as $Z$-factor contributions are negligible.

\subsection{How a singlet catalyzes strong transitions}

A relatively heavy singlet
with sufficiently strong portal interaction to
the doublet can enhance the transition strength by reducing
the effective quartic coupling for the doublet
\begin{align}
\label{eq:reduce:x:effect}
  x &\equiv \frac{\bar{\lambda}_{\phi,3}}{\bar{g}^2_3}
  \\ &\approx
  \frac{1}{\bar{g}^2_3} \bigg(
      \lambda_{\phi,3}
    + \frac{ \mu_{1,3} \lambda_{m,3}^{ } \mu_{m,3}^{ }}{2m_{s,3}^{4}}
    - \frac{\mu^2_{m,3}}{8 m^2_{s,3}}
    - \frac{\lambda^2_{m,3}}{32\pi m_{s,3}^{ }} \bigg)
    \,.
  \nonumber
\end{align}
Here,
we only kept terms that produce a significant effect,
i.e.\ tree-level cubic contributions
(for simplicity, we omitted terms $\propto\mu_3$) and
one-loop $Z_2$-symmetric contributions.%
\footnote{
  Subdominant effects come from temporal scalar and
  two-loop $Z_2$-symmetric singlet contributions
  in eq.~\eqref{eq:lam-softer} as well as
  one-loop cubic contributions
  in eq.~\eqref{eq:Gamma3-cubic}.
  Effects of two-loop temporal scalars
  in eq.~\eqref{eq:lam-softer} and
  other cubic contributions
  in eq.~\eqref{eq:Gamma3-cubic} are negligible.
  A similar organization of relative sizes of different contributions holds also for
  the mass parameter (and hence $y$);
  cf.\ eqs.~\eqref{eq:mass-softer} and \eqref{eq:Pi3-cubic}.
}
When the singlet tadpole $\mu_{1,3}$ is negative,
all singlet effects in eq.~\eqref{eq:reduce:x:effect}
come with opposite sign compared to $\lambda_{\phi,3}$, and hence
{\em reduce} $x$.
The presence of cubic couplings ease the realization of this effect significantly
as their effect appears already at tree-level.
In eq.~\eqref{eq:Gamma3-cubic},
we carefully inspected that higher-order corrections
preserve this effect.
In general,
two-loop effects to eq.~\eqref{eq:reduce:x:effect} tend to come with positive sign,
and hence
diminish the effect of reducing the effective doublet self-interaction.

To further understand the role of the singlet in a doublet-driven phase transition,
it is illuminating to consider the soft-to-softer EFT construction as follows.
The one-loop contribution to the effective potential from
the soft modes are
\begin{align}
\label{eq:veff-soft}
\Veff^{}(\varphi) &=
    \frac{1}{2} m^2_{\phi,3} \varphi^2
    + \frac{1}{4} \lambda_{\phi,3} \varphi^4
- \frac{1}{16\pi} g^3_3 \varphi^3
  \\ &
- \frac{1}{12\pi} \Big(
    3 (\mD^2 + \!\frac{1}{2}h_3 \varphi^2)^{\frac{3}{2}}
    + (m^2_{s,3} + \!\frac{1}{2}\lambda_{m,3} \varphi^2)^{\frac{3}{2}}
    \Big)
  \,,
  \nonumber
\end{align}
where we neglected the singlet cubic couplings for simplicity.
The last three terms are contributions of
the vector boson,
temporal scalar, and
the singlet,
with masses
\begin{align}
\mA^2 &= \frac{g^2_3 \varphi^2}{4}
\,,&
m_{\rmii{$A_0$}}^2 &= \mD^2 + \frac{1}{2}h_3 \varphi^2
\,,
\nn &&
m^2_{S} &= m^2_{s,3} + \frac{1}{2}\lambda_{m,3} \varphi^2
\,,
\end{align}
respectively.
The essence of the softer EFT construction is the mass hierarchy
$\mA^2 \ll m_{\rmii{$A_0$}}^2, m^2_S$
which holds for
field values towards the symmetric phase where the background field vanishes.
In the broken phase, expansions in
$h_3 \varphi^2/\mD^2$ and
$\lambda_{m,3} \varphi^2/m^2_{s,3}$
yield the one-loop contributions in
eqs.~\eqref{eq:lam-softer} and \eqref{eq:mass-softer}.
From eq.~\eqref{eq:lam-softer},
it is evident that
a larger portal interaction leads to
a smaller effective quartic coupling of the transitioning field, and
therefore smaller $x$, allowing to strengthen the transition.

For the temporal scalar,
this expansion is always sensible,
as the Debye mass $\mD$ is always soft as there is no
negative zero temperature contributions that could make it smaller.
For the singlet
$m^2_{s,3} \sim m^2_s + g^2 T^2$, and
the situation is different, since
$m^2_s$ could be negative.%
\footnote{
  In our numerical study in~\eqref{eq:BMB},
  we ensure that $m^2_s$ remains positive and sufficiently large to integrate out
  the singlet.
} 
In that case,
it is possible that
$m^2_{s,3} \ll \frac{1}{2}\lambda_{m,3} \varphi^2$ and
the singlet shall be treated on same footing as the spatial gauge field.
It is not integrated out together with the temporal scalar, but remains in
the final EFT and is treated in analogy to
the vector boson as described in appendix~\ref{sec:EFT:xpansion}.
In this case,
the cubic term in the LO potential becomes
\begin{align}
  - \frac{\varphi^3}{16\pi} \Big(
      \bar{g}^3_3
    + \frac{\sqrt{2}}{3} \lambda^{\frac32}_{m,3}
  \Big) \equiv
  - \frac{\varphi^3}{16\pi} \eta^3
  \,.
\end{align}
We observe, that
the singlet can significantly increase the height of
the potential barrier generated by the vector boson, and
therefore enhance the transition strength.
Or in other words, one can define a new, effective
$x_{\text{eff}} \equiv \bar{\lambda}_{\phi,3}/\eta^2$,
where
$\eta^2 \equiv (\bar{g}^3_3 + \frac{\sqrt{2}}{3} \lambda^{\frac32}_{m,3})^{\frac{2}{3}}$ and
larger
$\lambda_{m,3}$ lead to smaller $x_{\text{eff}}$, and hence stronger transitions.
In practical applications,
the portal coupling is significantly larger than
the gauge coupling.
In such a limit,
it is really the ratio
$\lambda_\phi/\lambda_m$ that controls the perturbative expansion.
Indeed, one should always organize
the perturbative power counting in terms of
the largest dimensionful coupling in a parent theory, in this case
$\lambda_m$.
The setup discussed in this paragraph was recently formulated for
general models in~\cite{Gould:2023ovu} up to and including N$^2$LO corrections.
With the technology presented in this \article{} together with
the companion article~\cite{Ekstedt:2023xxx},
such computations can be performed at maximal perturbative accuracy.

\subsection{Higher dimensional operators}
\label{sec:higher:dim}

Integrating out both
the soft temporal scalars and
the singlet, induces
the following contribution to the doublet higher dimensional operator%
\footnote{
  Naturally, several
  higher dimensional operators are generated at the same order.
  Such operators involve gauge fields and their effect for
  the phase-transition thermodynamics is often argued to be
  subdominant compared to the sextic scalar operator.
}
\begin{align}
\label{eq:dim6}
  \bigg[ c_{6}
  & + \frac{1}{192\pi} \Big(
      \frac{3 h^3_3}{\mD^3}
    + \frac{\lambda^3_{m,3}}{m^3_{s,3}}
  \Big)
  \nn&
  + \frac{\mu^2_{m,3}}{24} \Big(
      \frac{\lambda_{m,3}}{m^4_{s,3}}
    - \frac{\mu_{m,3} \mu_{3,3}}{m^6_{s,3}}
    \Big)
  \bigg] (\Phi^\dagger \Phi)^3
  \,.
\end{align}
Here,
$c_{6} \sim \mathcal{O}(g^6)$ encodes contributions from
the hard modes \cite{Schicho:2021gca}.
The remaining terms result from the soft modes, and
one-loop contributions scale as $\mathcal{O}(g^3/\pi)$ while
tree-level cubic terms are even larger, $\mathcal{O}(g^2)$.
In practice,
contributions from
the hard modes and
the temporal scalars produce a negligible effect, which is also the case for
singlet contributions,
provided that the singlet is sufficiently heavy.
We relegate a further quantitative analysis of
the EFT with the operator of eq.~\eqref{eq:dim6}
to future work.

We remark, that
the operator~\eqref{eq:dim6} {\em cannot} be neglected
when $\bar{\lambda}_{\phi,3}$ and hence $x$ become negative
in the vicinity of
the critical temperature.
This commonly occurs in many
theories
(cf.\ e.g.~\cite{Andersen:2017ika, Niemi:2018asa, Gould:2019qek}) and
happens also in our toy dark sector setup for
large (small) portal couplings (singlet masses).
For studies in which this higher dimensional operator is kept within
the 3D~EFT see~\cite{Croon:2020cgk,Ekstedt:2022ceo}
(cf.\ also \cite{Camargo-Molina:2021zgz}).

\subsection{Perturbative expansion}
\label{sec:pert:exp}

To capture
the thermal contributions enhanced in the IR,
we have utilized a chain of EFTs
starting from
the hard ($\pi T$) to
the soft ($g T$) to
the softer to
supersoft ($g^{\frac32}/\sqrt{\pi} T$) scale.
It is this high tower of effective theories above
the non-perturbative,
ultrasoft scale ($g^2/\pi T$) that brew the ultimate potion for resummations required to construct the pressure at high temperature.

In terms of
the formal power counting of a parent theory,
the different contributions to the pressure scale as
\begin{align}
\label{eq:pressure:expansion}
  p \sim T^4 \biggl(
  \underset{\text{``1-loop''}}{
    \fbox{$
    g^0 + g^2 + g^3
    \vphantom{g^{\frac{11}{2}}}
    $}}
  &+
  \underset{\text{``2-loop''}}{
    \fbox{$
    g^4 + g^{\frac{9}{2}\hspace{1.2mm}}
    \vphantom{g^{\frac{11}{2}}}
    $}}
  \nn
  &+
  \underset{\text{``3-loop''}}{
    \fbox{$
    g^5 + g^{\frac{11}{2}}
    \vphantom{g^{\frac{11}{2}}}
    $}}
  + \mathcal{O}(g^6)
  \biggr)
  \,.
\end{align}
The coupling orders contributing to
the different ``loop orders'' are convoluted%
\te
due to various resummations.
We comment on the following subtleties
\begin{itemize}
  \item[1-loop:]
    The orders
    $\mathcal{O}(g^0)$ and
    $\mathcal{O}(g^2)$
    only arise in the symmetric phase from hard modes.
    While the $\mathcal{O}(g^2)$-terms
    technically originate from two-loop diagrams,
    it is sensible to group them with
    the broken-phase LO $\mathcal{O}(g^3)$ soft contributions.
  \item[2-loop:]
    Terms of $\mathcal{O}(g^4)$ arise from various two-loop diagrams,
    except for hard contributions in the symmetric phase that appear at three-loop level. Terms of $\mathcal{O}(g^{\frac{9}{2}})$ appear at one-loop, but
    are be numerically close to $\mathcal{O}(g^4)$ terms.
  \item[3-loop:]
    Terms of
    $\mathcal{O}(g^5)$ and
    $\mathcal{O}(g^{\frac{11}{2}})$ require three-loop computations and are,
    similar to ``2-loop'', numerically almost identical
    in practice.
\end{itemize}
Since we do not perform strict expansions for
the pressure,
the speed of sound, and
$\alpha$, we have found the above grouping of different orders
practical for our results in
figs.~\ref{fig:pressure}--\ref{fig:alpha-beta}.
In turn, we do not truncate
the effective theory matching relations at
``1-loop'',
``2-loop'', or
``3-loop'', but work with
the full
relations in all cases.
As a consequence, many expressions at higher orders include
formally higher-order {\em tails} in
the full EFT spirit.

\begin{table}[t]
\begin{tabular}{|c|r|r|}
  \hline
  order &
  $\frac{p_\rmii{sym}}{p_{\rmii{sym},g^0\vphantom{_A}}}$\,&
  $\frac{\Delta \Veff}{\Delta V_{\rmii{eff},g^3}}$
  \\ \hline\hline
  $g^2$ &
  1.72\%
  &
  $-$
  \\
  $g^3$ &
  1.64\%
  &
  $-$
  \\
  $g^4$ &
  0.62\%
  &
  35.4\%
  \\
  $g^\frac{9}{2}$ &
  0.004\%
  &
  1.4\%
  \\
  $g^5$ &
  0.31\%
  &
  24.3\%
  \\
  $g^\frac{11}{2}$ &
  0.0001\%
  &
  1.1\%
  \\ \hline
\end{tabular}
  \caption{%
    Symmetric-phase pressure and free-energy difference
    relative to
    their respective leading order
    for benchmark point~\eqref{eq:BMA}.
    }
  \label{tab:p:corrections}
\end{table}

In~\eqref{eq:BMA},
for increasing orders in $g$ in eq.~\eqref{eq:pressure:expansion},
we find for
the symmetric-phase pressure that corrections relative to
its leading $\mathcal{O}(g^0)$ appear with
the importance listed in tab.~\ref{tab:p:corrections}.
The same table also lists the corrections for
the broken-phase pressure
$p_{\rmi{bro}} = p_{\rmi{sym}} - T \Delta \Veff$ and
free-energy difference,
relative to its leading contribution at $\mathcal{O}(g^3)$.
Since these numbers
depend on
the value of $T$ at which the pressure is evaluated,
they are taken to be tentative.
In particular,
we find that ``supersoft'' corrections with fractional powers are minute,
despite their apparent counting.
For this reason,
we group them together with integer power orders as in
eq.~\eqref{eq:pressure:expansion}.
Similar observations also hold for~\eqref{eq:BMB}
which we do not illustrate separately.

\bibliographystyle{utphys}
\bibliography{ref}

\end{document}